\documentclass[twocolumn,prl,amsmath,amssymb,superscriptaddress]{revtex4-1}
\usepackage{graphicx,bm,dsfont}
\usepackage{soul}
\usepackage{color}

\begin{document}

\title{Topologically Protected Quantum Coherence in a Superatom}

\author{Wei Nie}
\affiliation{Institute of Microelectronics, Tsinghua University, Beijing 100084, China}
\affiliation{Frontier Science Center for Quantum Information, Beijing, China}
\author{Z. H. Peng}
\affiliation{Key Laboratory of Low-Dimensional Quantum Structures and Quantum Control of Ministry of Education, Key Laboratory for Matter Microstructure and Function of Hunan Province, Department of Physics and Synergetic Innovation Center for Quantum Effects and Applications, Hunan Normal University, Changsha 410081, China}
\author{Franco Nori}
\affiliation{Theoretical Quantum Physics Laboratory, RIKEN Cluster for Pioneering Research, Wako-shi, Saitama 351-0198, Japan}
\affiliation{Physics Department, The University of Michigan, Ann Arbor, Michigan 48109-1040, USA}
\author{Yu-xi Liu}\email{yuxiliu@mail.tsinghua.edu.cn}
\affiliation{Institute of Microelectronics, Tsinghua University, Beijing 100084, China}
\affiliation{Frontier Science Center for Quantum Information, Beijing, China}

\begin{abstract}
Exploring the properties and applications of topological quantum states is essential to better understand topological matter. Here, we theoretically study a quasi-one-dimensional topological atom array. In the low-energy regime, the atom array is equivalent to a topological superatom. Driving the superatom in a cavity, we study the interaction between light and topological quantum states. We find that the edge states exhibit topology-protected quantum coherence, which can be characterized from the photon transmission. This quantum coherence helps us to find a superradiance-subradiance transition, and we also study its finite-size scaling behavior. The superradiance-subradiance transition also exists in symmetry-breaking systems. More importantly, it is shown that the quantum coherence of the subradiant edge state is robust to random noises, allowing the superatom to work as a  topologically protected quantum memory. We suggest a relevant experiment with three-dimensional circuit QED. Our study may have applications in quantum computation and quantum optics based on topological edge states.
\end{abstract}

\maketitle

\textit{Introduction}.---One of the most striking achievements in modern physics is the discovery of topological materials. Also, novel forms of topological quantum states are pursued in both matter and light~\cite{RevModPhys.82.3045,RevModPhys.83.1057,Bliokh2015Spin,RevModPhys.91.015006}. These exotic states are protected by band gaps which can be closed via topological phase transitions~\cite{PhysRevLett.108.220401,PhysRevLett.108.255303,xiaopengli2013}. Topological quantum states have applications in many quantum technologies, e.g., topological qubits~\cite{Kitaev2001,PhysRevB.81.014505,Alicea2011,PhysRevLett.110.076401,2014You}, topological quantum channels~\cite{2011arXiv1110.3788Y,dlaska2017robust}, topological surface waves~\cite{Bliokh2015Quantum,Bliokh2019Topological}, and topological lasing~\cite{Stjean2017Lasing,Bahari2017Nonreciprocal,Harari2018Topological,Bandres2018Topological}. In topological many-body systems, owing to the peculiar geometry of edge states, driving a single atom could excite an edge state and generate a quantum nonlinearity for photons~\cite{Perczel2017Topological}. In the emerging field of topological quantum optics~\cite{Pan2015,Mivehvar2017,Perczel2017Topological,PhysRevLett.119.173901,2017arXiv171100478B}, the interaction between light and topological quantum states should be explored to better understand the properties of topological quantum matter.

Collective behavior in quantum many-body systems originates from quantum coherence~\cite{PhysRev.93.99}. In cavity QED, single-photon absorption is able to build many-body coherence among atoms, producing superradiance or subradiance~\cite{PhysRevLett.102.143601,PhysRevLett.115.243602,PhysRevLett.116.163604,PhysRevLett.117.073003,PhysRevLett.120.193601}. A superatom model is used to explain such collective phenomena~\cite{vuletic2006quantum} and has been realized via Rydberg blockade~\cite{PhysRevLett.99.163601,PhysRevX.7.041010}. Recent studies about topological matter show that single-atom quantum coherence can be protected by topology~\cite{Bahri2015Localization,PhysRevA.96.053858,Kemp2017Long,PhysRevLett.119.123601}. Indeed, topological protection makes nonlocal quasiparticles in the ground state manifold ideal candidates for realizing topological quantum computation~\cite{PhysRevLett.99.020503,RevModPhys.80.1083}. In particular, researchers have analyzed quantum coherence of Majorana zero modes in decoherence-free subspaces~\cite{Diehl2011Topology} and quantum manipulation of Majorana bound states via electron-photon interactions~\cite{PhysRevA.79.040303,PhysRevLett.109.257002,PhysRevLett.110.107006,PhysRevLett.118.126803}.

We consider a quasi-one-dimensional (1D) topological array of two-level atoms. In the low-energy regime, the atom array
has a ground state and a single-excitation subspace which has many bulk states and two edge states. The large gaps between edge states and bulk states in the single-excitation subspace help us to define a topological superatom, which consists of a ground state and two edge states. The typical features of edge states make them experimentally measurable in various topological systems~\cite{Konig2007,Hsieh2009,Wang2009,Hafezi2011,Chang2013}. Here, we study edge states via light-matter interactions, from which topology-protected quantum coherence is found. Superconducting quantum circuits have applications in quantum computation and microwave photonics~\cite{Buluta,Gu2017Microwave}. The recent development of quantum chip technologies makes it possible to address qubit arrays, e.g., via 3D integration~\cite{PhysRevApplied.6.044010,QiangLiu2017,Rosenberg2017,Dunsworth2018}. For concreteness, here we propose an experimental setup for studying topological mater in an integrated superconducting quantum chip.

\begin{figure}[t]
\includegraphics[width=8.5cm]{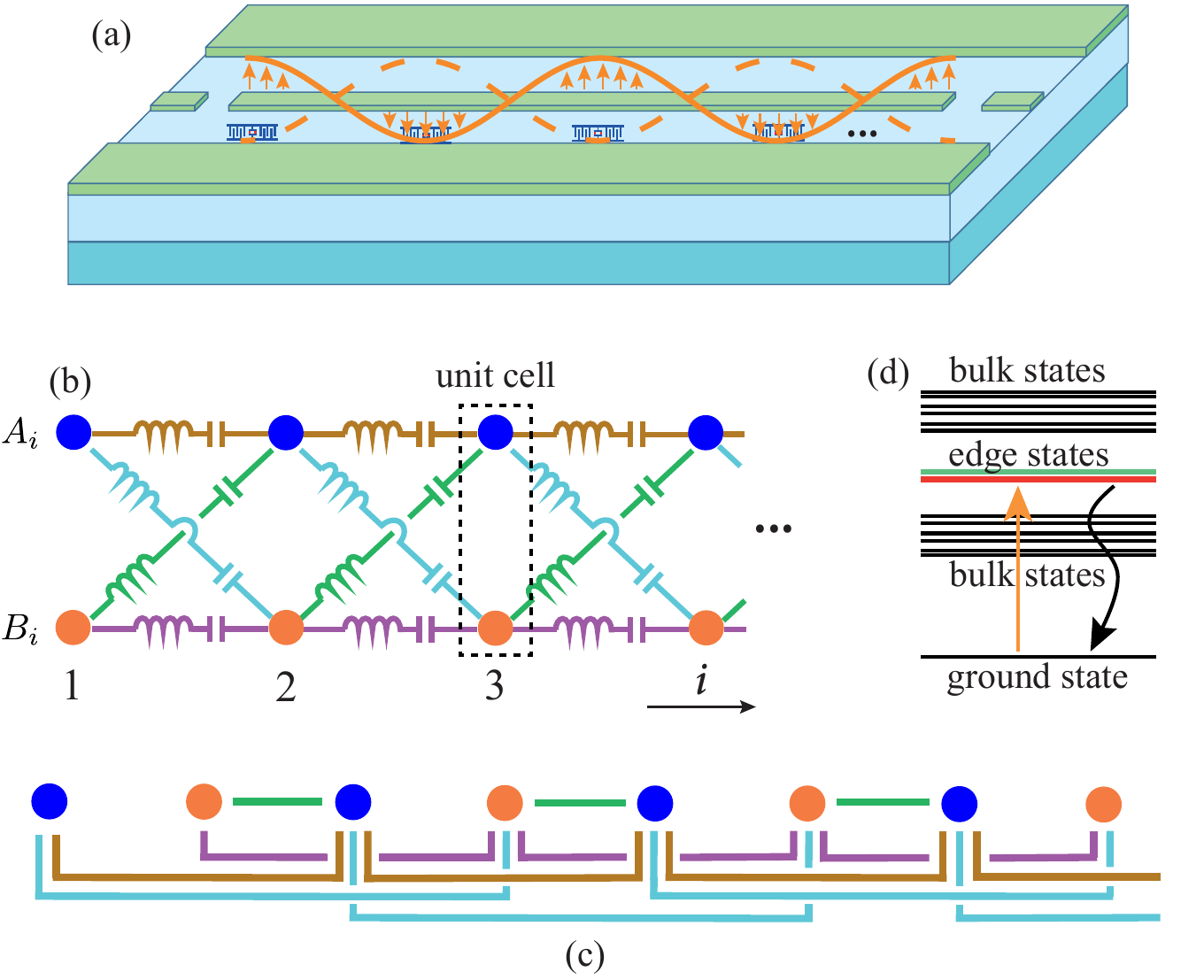}
\caption{(a) Schematic of a 3D circuit QED. The top layer contains a microwave transmission line resonator, which plays the role of cavity, coupled with an array of superconducting artificial atoms. On the bottom layer, superconducting coplanar waveguides are fabricated and coupled to the atoms on the top panel via interconnects in the middle dielectric layer (see Ref.~\cite{SupplementalMaterials} for details). (b) The atom array in (a) has internal interactions between neighboring unit cells. The atoms are coupled by resonators represented by \emph{LC} circuits. Blue and orange dots denote atoms $A$ and $B$ in unit cells. (c) Wiring of the coupling circuits, so a 1D atom array can be obtained and coupled to the transmission line resonator, as shown in (a). (d) Optically addressing edge states of the topological atom array.}\label{figure1}
\end{figure}

\textit{3D circuit QED with a topological atom array}.---Figure~\ref{figure1}(a) shows the schematic of a 3D circuit QED with multilayer fabrication process. The top layer consists of a transmission line resonator interacting with an artificial atom array. In the bottom layer, superconducting coplanar waveguides are fabricated (not shown). The atom array has a ladder configuration, as shown in Fig.~\ref{figure1}(b). The couplings between neighboring unit cells are realized by \emph{LC} resonators. Through 3D wiring, the ladder structure of the atom array can be reconfigured as a 1D array, as shown in Fig.~\ref{figure1}(c). The crossings between wires represent airbridges~\cite{Chen2014,Mukai2019,Rosenberg2019}. To show how the atoms are coupled, we first consider the interaction between atoms $A_1$ and $B_2$ in the first and second unit cells, respectively. In the rotating frame with the frequency of the coupler, the system Hamiltonian becomes $(\hbar=1)$
\begin{eqnarray}
H_{AB}=\sum_{\alpha=1A,2B} \Delta_{\alpha} \sigma_{\alpha}^+ \sigma_{\alpha}^- - g_{\alpha}( \sigma_{\alpha}^+ \hat{a}_1 + \hat{a}_1^{\dagger} \sigma_{\alpha}^-),
\end{eqnarray}
where $\Delta_{\alpha}$ and $g_{\alpha}$ are detunings and couplings between the atoms and the \emph{LC} resonator, respectively. Hereafter, we assume $\Delta_{1A}=\Delta_{2B}=\Delta$. Also, $\sigma_{1A}^+ = |A_1\rangle \langle \alpha_1 |$ and $\sigma_{2B}^+ = |B_2\rangle \langle \beta_2 |$ are the atomic operators where $|\alpha_1\rangle$ ($|A_1\rangle$) and $|\beta_2\rangle$ ($|B_2\rangle$) denote the ground (excited) states of atoms $A_1$ and $B_2$, respectively. And $\hat{a}_1$ ($\hat{a}^{\dagger}_1$) represents the annihilation (creation) operator of the resonator. When $g_{1A}, g_{2B} \ll |\Delta|$, by making a Schrieffer-Wolff transformation, we can obtain the effective Hamiltonian
\begin{eqnarray}
\tilde{H}_{AB}&=&\Big(\Delta + \frac{g_{1A}^2}{\Delta}\Big)\sigma_{1A}^+ \sigma_{1A}^- + \Big(\Delta + \frac{g_{2B}^2}{\Delta}\Big)\sigma_{2B}^+ \sigma_{2B}^- \nonumber \\
&& +\frac{g_{1A}g_{2B}}{\Delta} (\sigma_{1A}^+ \sigma_{2B}^- + \sigma_{2B}^+ \sigma_{1A}^-).
\label{intH}
\end{eqnarray}
The first and second terms contain Lamb shifts due to the virtual photons in the \emph{LC} resonator. The last term is the effective coupling between these two atoms, which can be realized in many quantum systems~\cite{PhysRevLett.85.2392,PhysRevLett.87.037902,Majer2007,Evans2018}. To couple two neighboring unit cells, we need four \emph{LC} resonators; each one producing a specific interaction. Based on this coupling scheme, an atom array can be obtained~\cite{SupplementalMaterials}.
\begin{figure}[b]
\includegraphics[width=8.5cm]{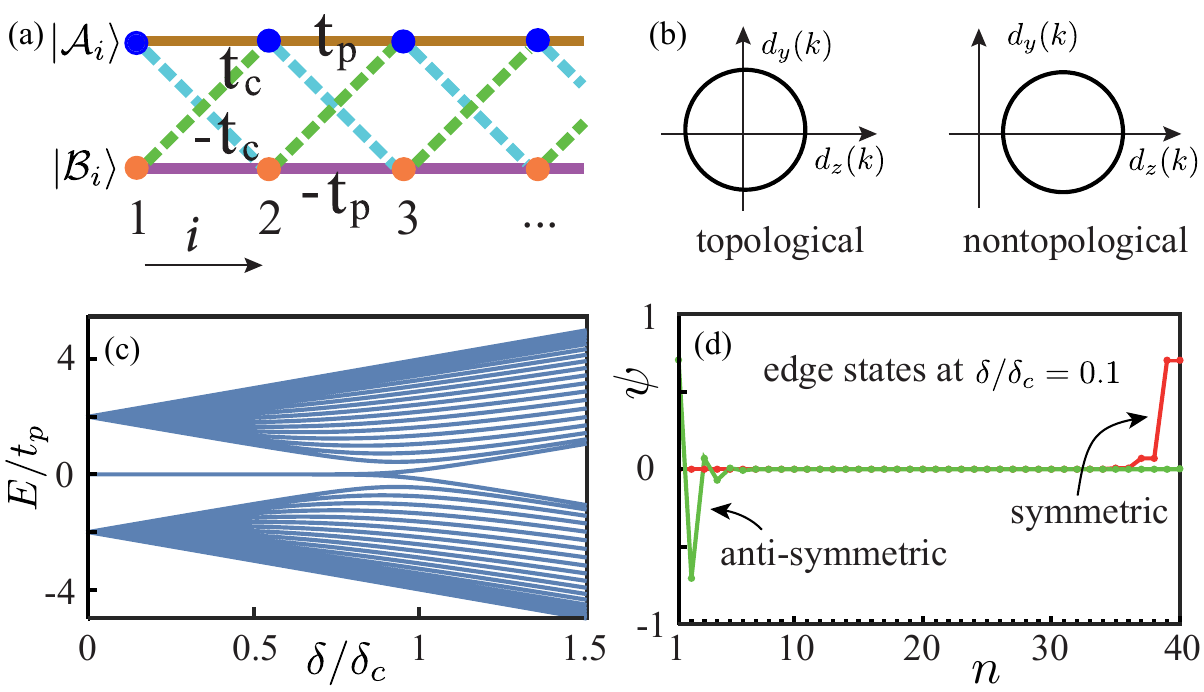}
\caption{(a) Lattice in the single-excitation subspace. Solid and dashed lines represent parallel and cross-couplings, respectively. (b) Topology of the lattice in the auxiliary space $[ d_y(k), d_z(k) ]$. The winding number for the topological phase is nontrivial. (c) Energy spectrum of the tight-binding lattice in (a). There are large gaps between edge states and bulk states. As $\delta$ changes across the critical point $\delta_c$, edge states undergo a transition to bulk states.  (d) Wave functions of edge states at $\delta=0.1\delta_c$. Here $n$ labels the positions of atoms in the array, and odd (even) number of $n$ corresponds to $|\mathcal{A}_{\frac{n+1}{2}}\rangle$ ($|\mathcal{B}_{\frac{n}{2}}\rangle$). The parameters in (c) and (d) are $t_c = t_p$ and the number of unit cells $N=20$.}\label{figure2}
\end{figure}

\textit{Topological superatom}.---The atomic interactions produced by exchanging virtual photons allow the study of many-body phenomena~\cite{PhysRevX.8.011002,PhysRevLett.120.050507,Norcia2018}. Using the airbridge wiring technique~\cite{Chen2014,Mukai2019,Rosenberg2019}, quantum networks of artificial atoms can be realized in superconducting quantum circuits. Considering the lattice in Fig.~\ref{figure1}(b), the effective Hamiltonian of the atom array can be written as
\begin{eqnarray}
\tilde{H}&=& \sum_{i=1}^N \delta (\sigma_{iA}^+ \sigma_{iA}^- - \sigma_{iB}^+ \sigma_{iB}^-) +  \sum_{i=1}^{N-1}\Big[ t_p (\sigma_{iA}^+ \sigma_{i+1A}^- \nonumber \\
&-& \sigma_{iB}^+ \sigma_{i+1B}^- ) - t_c (\sigma_{iA}^+ \sigma_{i+1B}^- - \sigma_{iB}^+ \sigma_{i+1A}^- ) + \mathrm{H.c.} \Big], \label{effH}
\end{eqnarray}
where $\delta$ is half of the effective energy splitting between two excited states $|A_i\rangle$ and $|B_i\rangle$ of atoms $A$ and $B$ in the \emph{i}th unit cell; $t_p$ and $t_c$ are, respectively, the parallel and cross-couplings~\cite{SupplementalMaterials}. To better see the physical picture of Eq.~(\ref{effH}), we can rewrite it in the single-excitation subspace $\{|\mathcal{A}_i\rangle, |\mathcal{B}_i\rangle \}$, with $|\mathcal{A}_i\rangle = \sigma_{iA}^+ |G\rangle$ and $|\mathcal{B}_i\rangle = \sigma_{iB}^+ |G\rangle$ (here $|G\rangle = |\alpha_1\beta_1\alpha_2\beta_2\cdots\rangle$), which represents a lattice as shown in Fig.~\ref{figure2}(a). After making Fourier transforms to the vectors $|\mathcal{A}_i\rangle$ and $|\mathcal{B}_i\rangle$, Eq.~(\ref{effH}) can be written in crystal momentum space as $\bar{H}(k) = \sum_{k} \Psi_k^{\dagger} h(k) \Psi_k$, with $\Psi_k^{\dagger} = (|\mathcal{A}_k \rangle,|\mathcal{B}_k \rangle)$, and
\begin{equation}
h(k) = d_y(k) \sigma_y + d_z(k) \sigma_z.
\end{equation}
Here, $d_y(k) = 2 t_c \sin k$ and $d_z(k) = \delta + 2 t_p \cos k$. The system is protected by chiral symmetry~\cite{ryu2002}, i.e., $\sigma_x h(k) \sigma_x = -h(k)$, as well as particle-hole and time-reversal symmetries, and belongs to the BDI class~\cite{PhysRevB.78.195125}. The topological nature can be extracted from the winding number~\cite{PhysRevLett.115.177204,TaoLiu2018}, defined in the auxiliary space $[ d_y(k), d_z(k) ]$, as shown in Fig.~\ref{figure2}(b). When $-\delta_c < \delta < \delta_c$, with $\delta_c = 2 |t_p|$, the system is in a topological insulating phase with nontrivial winding number. As $|\delta|$ increases and becomes larger than $\delta_c$, a normal insulator is obtained for zero winding number.

From the edge-bulk correspondence, it is known that the topological phase supports edge states for open boundary conditions. The energy spectrum of the atom array in the single-excitation subspace is shown in Fig.~\ref{figure2}(c).  Zero modes for $|\delta| < \delta_c$ represent edge states. The edge states localized at the left and right boundaries are
\begin{eqnarray}
\psi_L &=& [\mathcal{N}_L^-]^{-\frac{1}{2}}\sum_i \left[(\lambda_{-,1})^i - (\lambda_{-,2})^i\right] \phi^{(i)}_{-}, \label{edgewave1}  \\
\psi_R &=& [\mathcal{N}_R^+]^{-\frac{1}{2}}\sum_i \left[(\lambda_{-,1})^{N+1-i} - (\lambda_{-,2})^{N+1-i}\right] \phi^{(i)}_{+}, \label{edgewave2}
\end{eqnarray}
where $\mathcal{N}_L^-$ and $\mathcal{N}_R^+$ are the renormalization factors and $\lambda_{-,l} = \big[\delta + (-1)^{l-1} (\delta^2 - 4 t_p^2 + 4 t_c^2)^{1/2}\big]/ (-2t_c- 2t_p)$ (with $l=1,2$)~\cite{PhysRevLett.110.076401,Konig08}, $\phi^{(i)}_{\pm}=|\mathcal{A}_i\rangle\pm|\mathcal{B}_i\rangle$. From the edge states, we can find several features. First, the left and right edge states are polarized with antisymmetric and symmetric superpositions of $|\mathcal{A}_i\rangle$ and $|\mathcal{B}_i\rangle$, respectively. Second, the edge states are exponentially localized in the boundaries, as shown in Fig.~\ref{figure2}(d). These properties are helpful for manipulating edge states. The above edge states occur when $|\lambda_{-,l}|<1$. The case $|\lambda_{-,l}|>1$ has oppositely polarized edge states~\cite{SupplementalMaterials}. From the spectrum, we can find that the edge states have large energy gaps with bulk states. Therefore, a topological superatom with a $V$-shaped three-level structure~\cite{PhysRevB.92.020515,PhysRevA.94.033829}, which consists of a ground state and two edge states, can be modeled to characterize the atom array in its low-energy regime.

\textit{Optically probing edge states}.---Generally speaking, it is challenging to selectively drive quantum many-body states in large-scale systems. However, owing to specific properties of  the edge states analyzed above, one can realize interactions between light and edge states. As shown in Fig.~\ref{figure1}(a), the atom array can be driven by a single-mode cavity field. The Hamiltonian of the cavity field with external driving is $H_c = \Delta_c \hat{f}^{\dagger} \hat{f} + i \eta(\hat{f}^{\dagger} - \hat{f})$,
where $\Delta_c = \omega_c -\omega_l$, $\hat{f}$ ($\hat{f}^{\dagger}$) is annihilation (creation) operator of the cavity field, $\eta$ is the pumping strength, and $\omega_c$ and $\omega_l$ are the frequencies of the cavity and driving fields, respectively. The Hamiltonian describing the couplings between the cavity field and the atom array is
$H_{I} = \sum_{i} ( \xi_{iA} \hat{f} \sigma_{iA}^+ + \xi_{iB} \hat{f} \sigma_{iB}^+ + \mathrm{H.c.})$. We consider the resonant driving of edge states, and the large gaps between edge states and bulk states prevent bulk states from being excited. The dynamics of the many-body system is described by the master equation
$\dot{\rho} = i [\rho,H_{\mathrm{tot}}] + \mathcal{L}_c[\rho] + \mathcal{L}_a[\rho]$,
with the total Hamiltonian $H_{\mathrm{tot}} =\tilde{H}+ H_c + H_I$, and dissipation terms for the cavity $\mathcal{L}_c[\rho] = \kappa(2 \hat{f} \rho \hat{f}^{\dagger}-\hat{f}^{\dagger} \hat{f} \rho - \rho \hat{f}^{\dagger} \hat{f})$ and atom array $\mathcal{L}_a[\rho] = \sum_{i,\mu,\nu} \gamma_{\mu\nu} (2\sigma_{i\mu}^- \rho \sigma_{i\nu}^+ - \sigma_{i\mu}^+ \sigma_{i\nu}^- \rho - \rho \sigma_{i\mu}^+ \sigma_{i\nu}^-)$. Here, $\kappa$ is the decay rate of the cavity, and $\gamma_{\mu\nu}$ the decay rates of the atoms~\cite{Nathan2018}. Specifically, $\gamma_{AA},\gamma_{BB}$ are the decay rates of atoms $A_i$ and $B_i$, respectively. For simplicity, we write $\gamma_{AA}=\gamma_{BB}=\gamma$. The correlated decays $\gamma_{AB}$ and $\gamma_{BA}$ between atoms $A_i$ and $B_i$ play fundamental roles in many quantum optical effects~\cite{Agarwal1974,PhysRevLett.77.3995,PhysRevLett.76.388,PhysRevLett.84.5500,PhysRevLett.100.043601,PhysRevLett.101.153601}. The symmetric correlated decays, i.e., $\gamma_{AB}=\gamma_{BA}$, can be realized by coupling two atoms to a waveguide~\cite{PhysRevLett.106.020501,vanLoo2013,Mohammad2018}. In the 3D integrated circuits [see Fig.~\ref{figure1}(a)], the artificial atoms are coupled to superconducting coplanar waveguides via interconnects~\cite{SupplementalMaterials}.

\begin{figure}[b]
\includegraphics[width=8.5cm]{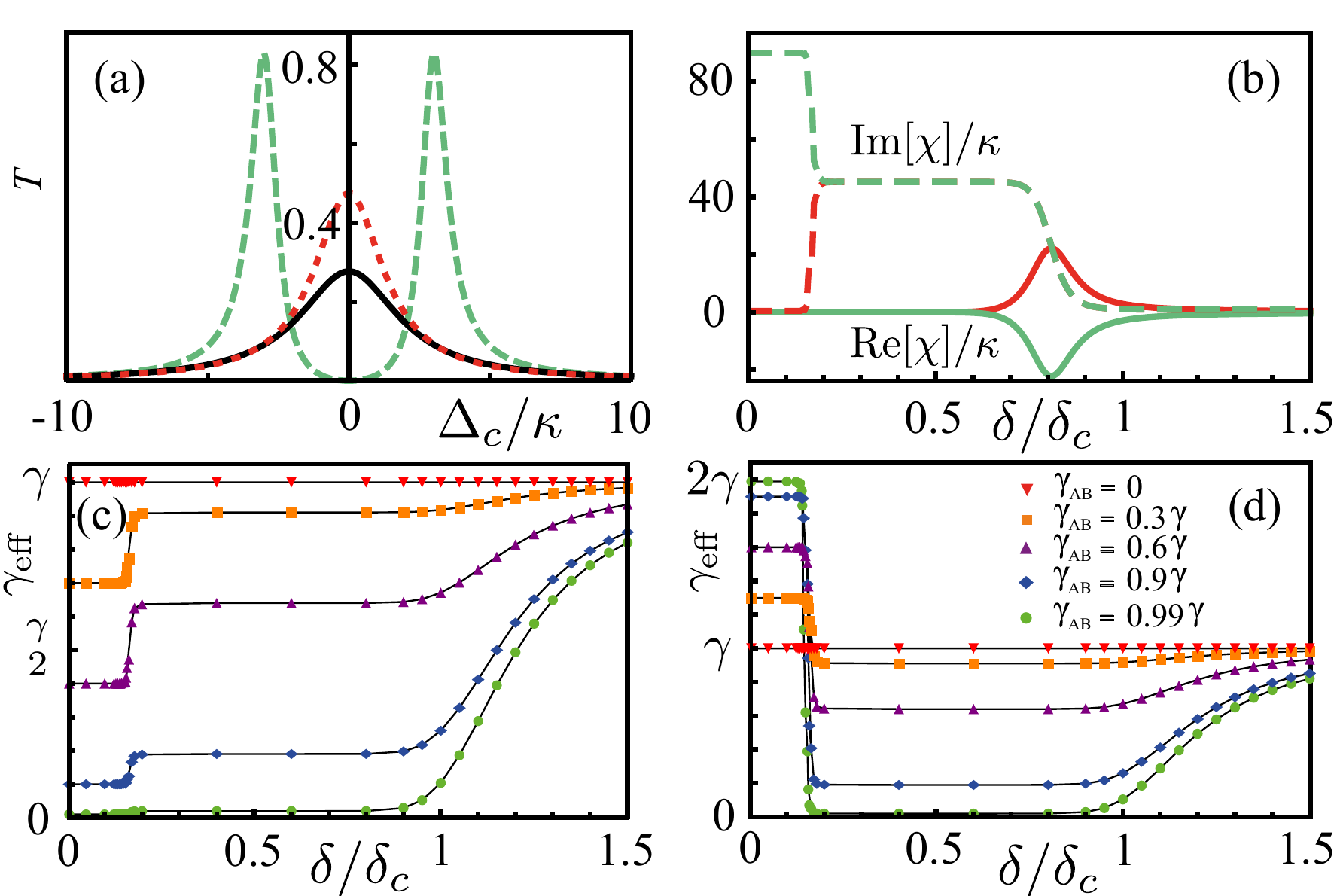}
\caption{(a) Transmission of light through the left (green-dashed) and right (red-dotted) edge states with $\gamma_{AB}=0.99\gamma$. The black curve represents the transmission for both left and right edge states with $\gamma_{AB}=0$. Here we consider $\delta=0.1\delta_c$. (b) Real (solid) and imaginary (dashed) parts of the rescaled susceptibility $\chi$ by $\kappa$ for the left (green) and right (red) edge states with $\gamma_{AB}=0.99\gamma$. $\mathrm{Im}[\chi]$ shows the edge-bulk transition in a finite lattice. (c,d) Variations of coherence when the system is changed from the topological to the nontopological regime. The effective decays of (c) and (d) in the unhybridized regime $\delta<0.15\delta_c$ correspond to the left and right edge states, respectively. The $\gamma_{AB}$ used in (c) is the same as that in (d). Other parameters for these figures: $t_c = t_p, \gamma=10\kappa, N=20$.}\label{figure3}
\end{figure}

In the low-excitation limit, the dynamic equations of the system are
\begin{eqnarray}
\big\langle \frac{d}{dt} \hat{f}\big\rangle &=& -(\kappa + i \Delta_c) \langle \hat{f} \rangle -i \bm{\Xi}^{\mathrm{T}} \langle \bm{\sigma} \rangle + \eta, \label{m1}  \\
\big\langle \frac{d}{dt}\bm{\sigma} \big\rangle &=& -i (\bm{\Delta} + \bm{D} -i \bm{\Gamma}) \langle \bm{\sigma} \rangle -i \bm{\Xi} \langle \hat{f} \rangle, \label{m2}
\end{eqnarray}
where $\bm{\Xi}$ is the coupling vector between cavity field and atoms. Also, $\langle \bm{\sigma} \rangle=(\langle \sigma_{1A}^- \rangle, \langle \sigma_{1B}^- \rangle, \langle \sigma_{2A}^- \rangle, \langle \sigma_{2B}^- \rangle, \cdots)^{\mathrm{T}}$, $\bm{\Delta}=\mathrm{Diag}(\delta,-\delta,\delta,-\delta,\cdots)$; while $\bm{D}$ and $\bm{\Gamma}$ denote the couplings and dissipations in the atom array~\cite{SupplementalMaterials}. From Eqs.~(\ref{m1}) and (\ref{m2}), the transmission can be formulated as
\begin{equation}
T=\Big|\frac{\kappa}{\kappa + i \Delta_c -i \chi}\Big|^2, \label{eqtr}
\end{equation}
where the susceptibility is $\chi=\bm{\Xi}^\intercal (\bm{\Delta} + \bm{D} -i \bm{\Gamma})^{-1} \bm{\Xi}$. When the cavity field is resonant with the superatom and the coupling parameters are appropriately chosen, only the edge state can be driven. It is known that in cavity QED with a single atom, the photon transmission exhibits radiation properties of the atom~\cite{McKeever2003,Astafiev2010Resonance}. For the topological superatom here, properties of edge states can be explored. Figure~\ref{figure3}(a) presents the transmission corresponding to the left and right edge states for $\delta=0.1\delta_c$. As the correlated decay $\gamma_{AB}$ increases, the transmission for the left edge state at resonance decreases. However, for the right edge state, the transmission is enhanced accordingly. The cavity decay $\kappa$ plays an important role in the transmission. Here we consider the cavity with low decay, i.e., $\kappa=0.1\gamma$. The large cavity decay is also studied~\cite{SupplementalMaterials}. Light transmission is versatile in detecting topological states~\cite{PhysRevA.81.033622,Rechtsmann2013,xiao2014,mei2015}. Figure~\ref{figure3}(b) shows the rescaled $\mathrm{Re}[\chi]$ (solid line) and $\mathrm{Im}[\chi]$ (dashed line) for both the left (green) and right (red) edge states. At $\delta=0.1\delta_c$, as studied in Fig.~\ref{figure3}(a), $\mathrm{Re}[\chi]$ is zero [see Fig.~\ref{figure3}(b)]; therefore, the transmission at resonance is $T_{\mathrm{res}}=1/(1+\mathrm{Im}[\chi]/\kappa)^2$. For the left (right) edge state, $\mathrm{Im}[\chi]/\kappa$ is $90$ ($0.45$) and $T_{\mathrm{res}}$ is about $0$ ($0.48$), for the given parameters. Figure~\ref{figure3}(b) shows two regimes with different values of $\mathrm{Im}[\chi]$ in the topological phase, produced by a finite-size topological phase transition. The detailed physics will be discussed below.

\begin{figure}[t]
\begin{center}
\includegraphics[width=8.5cm]{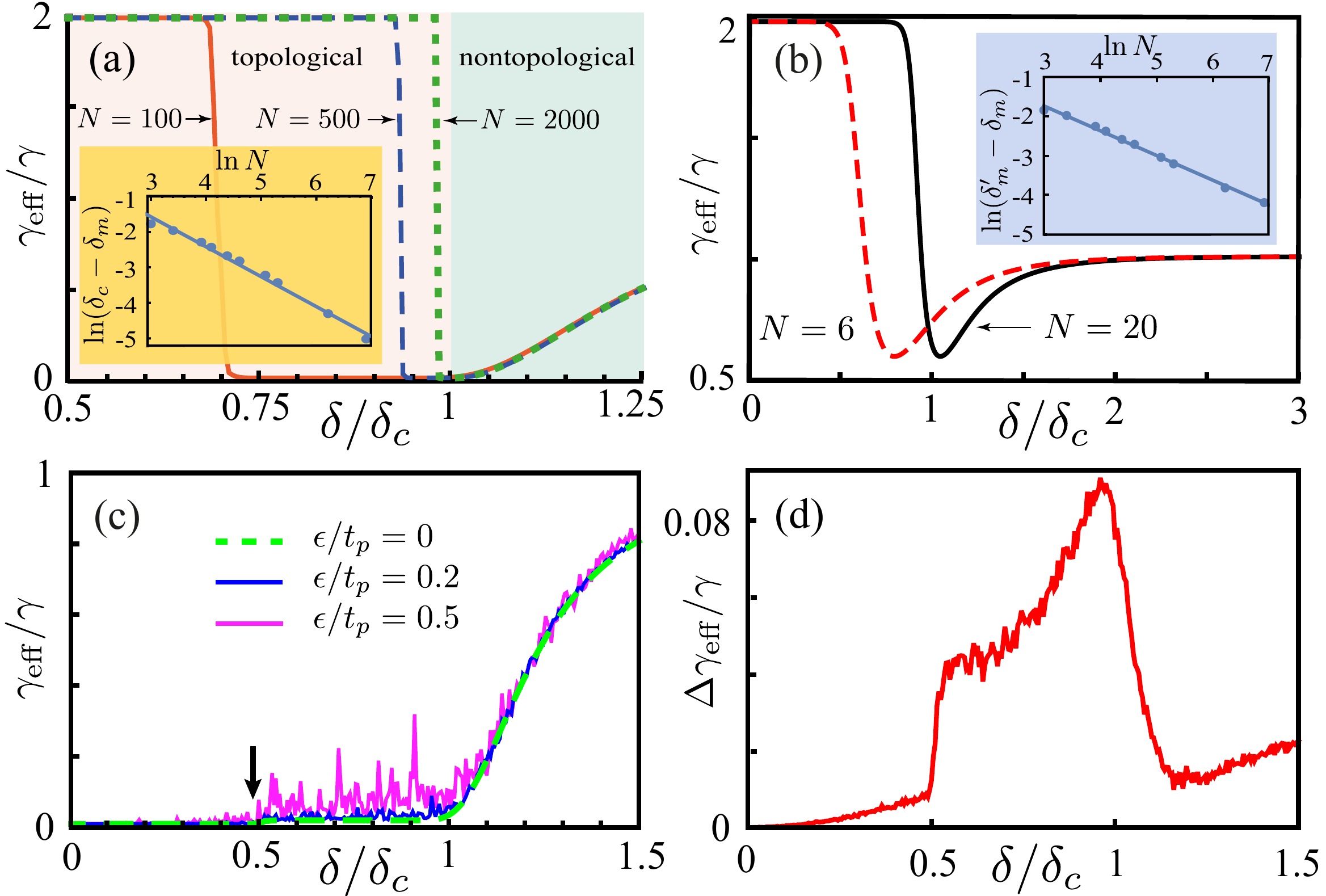}
\end{center}
\caption{(a) Superradiance-subradiance transition with different lattice lengths. The inset shows the finite-size scaling of SST for $\gamma_{AB}=0.99\gamma$, where $\delta_m$ is defined by $\gamma_{\mathrm{eff}}(\delta_m)=\gamma$. (b) Effect of symmetry breaking resulting from waveguide-induced interactions between atoms, with $\gamma_{AB}=0.97\gamma$ and the interactions between atoms in the same unit cells $g_{AB}=0.1\gamma$. The inset shows the finite-size scaling behavior of $\mathrm{ln}(\delta_m'-\delta_m)$, where $\delta_m'$ indicates the SST in the symmetry-breaking case. (c) The effect of disorder in atomic frequencies, with $\gamma_{AB}=0.99\gamma$ and $N=50$. The arrow indicates the position $\delta=\delta_m$, where SST takes place. (d) The difference between averaged $\gamma_{\mathrm{eff}}$ with disorder ($\epsilon/t_p=0.5$) and $\gamma_{\mathrm{eff}}$ without disorder. Other parameters for these figures: $t_c = t_p, \gamma=10\kappa$.}\label{figure4}
\end{figure}

\textit{Quantum coherence of topological superatom}.---From the susceptibility, we can obtain the effective decay, $\gamma_{\mathrm{eff}}=-\mathrm{Im}(\bm{\Xi}^\intercal \bm{\Xi}/\chi)$ as in Refs.~\cite{PhysRevLett.117.210503,PhysRevLett.119.093601,Albrecht1}, of the topological superatom. Based on the coupling between light and edge states, we explore the quantum coherence, which can be inferred from $\gamma_{\mathrm{eff}}$~\cite{PhysRevLett.121.073602}, in a topological superatom. We find that all the eigenmodes have the same coherence for $\gamma_{AB}=0$. However, as shown in Figs.~\ref{figure3}(c) and \ref{figure3}(d), the coherence properties of the superatom vary when $\delta$ is changed from the topological ($\delta<\delta_c$) to the nontopological ($\delta>\delta_c$) regime for nonzero $\gamma_{AB}$. In the topological regime, as shown in Fig.~\ref{figure3}(d), we find the superradiance-subradiance transition (SST) by defining $\gamma_{\mathrm{eff}}(\delta_m) = \gamma$, where the transition point $\delta_m=0.15\delta_c$ for given parameters characterizes the hybridization of the edge states. In the unhybridized regime $\delta<\delta_m$, the left edge state is subradiant $(\gamma_{\mathrm{eff},L}=\gamma - \gamma_{AB} < \gamma)$, and the right edge state is superradiant $(\gamma_{\mathrm{eff},R}=\gamma + \gamma_{AB} > \gamma)$. But, the hybridized edge states in the regime $\delta_m<\delta<\delta_c$ are subradiant, as shown in Figs.~\ref{figure3}(c) and \ref{figure3}(d).

In Fig.~\ref{figure4}(a), we further study the SST for different sizes of atom arrays. The inset presents the finite-size scaling behavior between the SST and the topological phase transition. The effective decay starts to increase after the topological phase transition. The symmetries in the atom array can be broken when waveguides induce interactions between atoms. In this scenario, the degeneracy of edge states is shifted, while the polarizations of edge states are preserved~\cite{SupplementalMaterials}. The SST is still found, as shown in Fig.~\ref{figure4}(b). The inset shows the finite-size scaling behavior between SSTs for symmetry-breaking and symmetry-preserving cases. The shift of the SST produced by symmetry-breaking interactions depends on system's size. In Fig.~\ref{figure4}(c), we study the disorder effect of atomic frequencies $\omega_{i\alpha}+\epsilon_{i\alpha}$ ($\alpha=A,B$), where the $\epsilon_{i\alpha}$ are randomly distributed $\epsilon_{i\alpha}\in[-\epsilon,\epsilon]$. Here, $\epsilon$ represents the strength of the disorder. The quantum coherence of the subradiant edge state without hybridization ($\delta<\delta_m$) is robust to random noise. However, the noise induces decoherence for hybridized edge states ($\delta_m<\delta<\delta_c$). In Fig.~\ref{figure4}(d), we characterize the disorder-induced decoherence by $\Delta\gamma_{\mathrm{eff}}=\overline{\gamma_{\mathrm{eff}}}-\gamma_{\mathrm{eff}}$, where $\overline{\gamma_{\mathrm{eff}}}$ is the averaged effective decay of the disordered systems. The unhybridized subradiant edge state is indeed robust to noises, compared with the hybridized edge states and bulk states. It can be used for quantum memory.

\textit{Discussions and conclusions}.---Recently, a ladder array with $24$ superconducting artificial atoms has been experimentally demonstrated~\cite{PhysRevLett.123.050502}. We find that even in small-size topological atom arrays realizable in current experiments, the collective edge states studied here can be observed. For example, when the number of unit cells is $N=6$, the edge states are localized and allow for optical measurements~\cite{SupplementalMaterials}. The correlated decay $\gamma_{AB}=0.99\gamma$ we considered in this work means a Purcell factor $\sim 100$, which has been realized in superconducting quantum circuits~\cite{Mohammad2018}.  Moreover, by considering the fluctuations of atomic frequencies and interactions, we find that the quantum coherence of edge states can be robust to random noises~\cite{SupplementalMaterials}.

In summary, we propose a quantum optical method to study topological matter. Owing to the large gaps between edge states and bulk states, a topological superatom is able to characterize the atom array in the low-energy regime. To optically drive the superatom, the unique properties of edge states (i.e., topology-protected polarization and boundary localization) are utilized. From the photon transmission, we find topology-protected quantum coherence distributed in the superatom. The topological superradiance and subradiance found here have important applications. When the symmetries in the system are preserved, the SST has a finite-size scaling relation with the topological critical point. This means that quantum coherence may provide an alternative way to characterize topological phases~\cite{PhysRevLett.119.250401,PhysRevLett.120.250501}. The SST is still found in symmetry-breaking systems, and the symmetry-breaking-induced shift of the SST depends on the system size. We study the effect of disorder on the system parameters and find that the quantum coherence of the unhybridized subradiant edge state is robust to random noises. Therefore the superatom can be used as a topology-protected quantum memory. We hope that this proposal can be experimentally realized by a circuit-QED system.

\begin{acknowledgments}
The authors thank Xiong-Jun Liu, Jiang Zhang, and Ying Li for helpful discussions, and acknowledge Tao Liu and Yu-Ran Zhang for a critical reading. Y.X.L. is supported by the Key-Area Research and Development Program of GuangDong Province under Grant No. 2018B030326001, the National Basic Research Program (973) of China under Grant No. 2017YFA0304304, and NSFC under Grant No. 11874037. W.N. acknowledges the Tsinghua University Postdoctoral Support Program. Z.H.P. is supported by NSFC under Grant No. 61833010 and Hunan Province Science and Technology Innovation Platform and Talent Plan (Excellent Talent Award) under Grant No. 2017XK2021. F.N. is supported in part by the MURI Center for Dynamic Magneto-Optics via the Air Force Office of Scientific Research (AFOSR) Grant No. FA9550-14-1-0040,
Army Research Office (ARO) Grant No. W911NF-18-1-0358, Asian Office of Aerospace Research and Development (AOARD) Grant No. FA2386-18-1-4045, Japan Science and Technology Agency (JST) through the Q-LEAP program and CREST Grant No. JPMJCR1676, the Japan Society for the Promotion of Science (JSPS) through the JSPS-RFBR Grant No. 17-52-50023 and JSPS-FWO Grant No. VS.059.18N, the RIKEN-AIST Challenge Research Fund, the Foundational Questions Institute (FQXi), and the NTT Physics and Informatics (NTT-PHI) Laboratory.
\end{acknowledgments}

\pagebreak
\clearpage

\onecolumngrid
\flushbottom

\begin{center}
	\section{Topologically Protected Quantum Coherence in a Superatom -- Supplemental Material}
\end{center}

\setcounter{equation}{0}
\setcounter{figure}{0}
\renewcommand{\thefigure}{S\arabic{figure}}
\renewcommand{\theequation}{S\arabic{equation}}
\setcounter{secnumdepth}{3}

\makeatletter
\def\@hangfrom@section#1#2#3{\@hangfrom{#1#2#3}}
\makeatother

%\maketitle

\section{3D integrated superconducting quantum circuits}
\begin{figure}[!htbp]
\includegraphics[width=14cm]{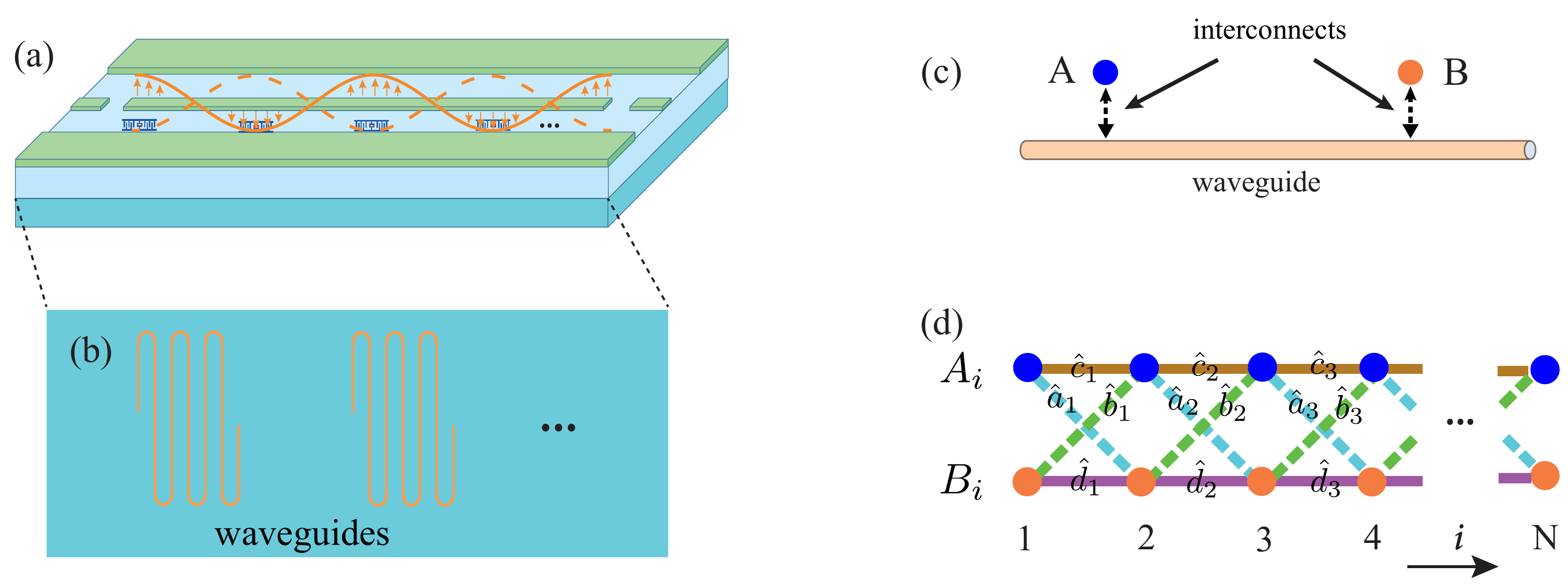}
\caption{(a) Schematic of 3D circuit QED with a topological atom array. The atoms in the array interact with their neighboring atoms via \emph{LC} resonators, as shown in Fig.1(b) in the main text. (b) The bottom layer with superconducting coplanar waveguides. Each waveguide couples to a unit cell on the top layer. (c) The atoms $A$ and $B$ in a unit cell couple to a waveguide via interconnects in the middle layer. (d) Resonator-mediated atom array (see Fig.1(b) in the main text). Here the operators $\hat{\mu}_j$, with $\mu=a,b,c,d$ and $j\in [1,N-1]$, correspond to resonator modes. The index $i$ labels the unit cell of the lattice.}\label{figS1}
\end{figure}

We consider 3D integrated superconducting quantum circuits~\cite{PhysRevApplied.6.044010,Rosenberg2017SM} to simulate and detect many-body systems. In quantum computation, two-dimensional arrays require multi-layer wiring~\cite{Dunsworth2018SM,Mukai2019SM}. In Fig.~\ref{figS1}(a), we show the 3D circuit QED with an atom array. The top layer contains a transmission line resonator and an atom array. The atoms in the array are coupled by \emph{LC} resonators (not shown). The superconducting coplanar waveguides are fabricated on the bottom layer, as shown in Fig.~\ref{figS1}(b). The atoms can be coupled to the waveguides via vertical interconnects~\cite{PhysRevApplied.6.044010,Rosenberg2017SM}. Here, we consider that these two atoms in the same unit cells are coupled to the same waveguides. The coupling of two atoms to a waveguide is presented in Fig.~\ref{figS1}(c). We assume that atoms $A$ and $B$ have the same frequency $\omega_0$. The effective coupling and correlated decay of these two atoms are~\cite{Tudela2011,Mohammad},
\begin{equation}
g_{AB} = \frac{\gamma_0}{2} \sin \Big(\frac{2\pi d_{AB}}{\lambda_0}\Big), \quad\quad
\gamma_{AB} = \gamma_0 \cos \Big(\frac{2\pi d_{AB}}{\lambda_0}\Big), \label{Eq_gAB}
\end{equation}
respectively. Here, $\gamma_0$ is the decay rate of the atoms to the waveguide, $\lambda_0=2\pi c/\omega_0$, and $d_{AB}$ is the distance between atoms $A$ and $B$ along the waveguide. As the positions of the atoms are properly chosen, e.g., $d_{AB}=m \lambda_0$ ($m$ is an integer), the interaction between these two atoms can be zero, but the correlated decay of the two atoms is maximum~\cite{Mohammad}. In Fig.~\ref{figS1}(d), we show the atom array coupled by \emph{LC} resonators. The resonator modes are represented by operators $\hat{a}_j,\hat{b}_j,\hat{c}_j,\hat{d}_j$ with $j\in [1,N-1]$.

\begin{figure}[t]
\includegraphics[width=12cm]{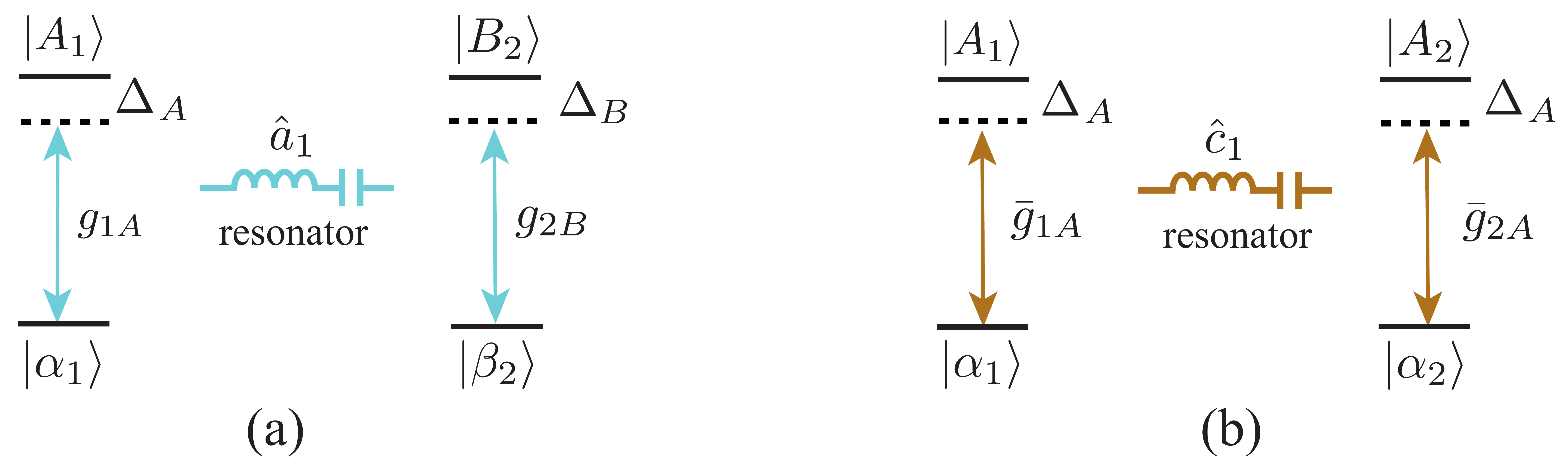}
\caption{Single-resonator-mediated two atoms. (a) The coupler $\hat{a}_1$ mediates the interaction between atoms $A_1$ and $B_2$. (b) The two atoms $A_1$ and $A_2$ are mediated by a resonator $\hat{c}_1$. }\label{figS2}
\end{figure}

\subsection{Single-resonator coupled two atoms}

As an example, we consider the resonator-mediated interaction between atom $A$ and and atom $B$ in the first and second unit cells [shown in Fig.~\ref{figS2}(a)], respectively. The Hamiltonian is
\begin{eqnarray}
H_{AB} =&& \omega_{a_1} \hat{a}^{\dagger}_1\hat{a}_1 + \omega_{1A} \sigma_{1A}^+ \sigma_{1A}^-  + \omega_{2B} \sigma_{2B}^+ \sigma_{2B}^- -g_{1A} (\hat{a}^{\dagger}_1 \sigma_{1A}^-  + \sigma_{1A}^+\hat{a}_1)   - g_{2B} (\hat{a}^{\dagger}_1 \sigma_{2B}^- + \sigma_{2B}^+ \hat{a}_1), \label{H2qubit}
\end{eqnarray}
with $\omega_{1A}=\omega_{2B}=\omega_0$. Here, $\hat{a}_1$ and $\hat{a}^{\dagger}_1$ represent the annihilation and creation operators of the \emph{LC} resonator mode that couples to the $A_1$ and $B_2$ atoms. The operators for atoms $A$ and $B$ are $\sigma_{1A}^+ = |A_1\rangle \langle \alpha_1 |$ and  $\sigma_{2B}^+ = |B_2\rangle \langle \beta_2 |$. We use $|\alpha_i\rangle$ and $|\beta_i\rangle$ to represent the ground states of the $A$ and $B$ atoms in the \emph{i}th unit cell. We use $g_{i\mu}$ ($\mu=A,B$) to denote the resonator-atom couplings [see Fig.~\ref{figS2}(a)]. In Eq.~(\ref{H2qubit}), the total number of excitations is conserved. Therefore, we can rewrite the Hamiltonian in a rotating frame with $H_{\mathrm{rot}}=\omega_{a_1} (\hat{a}^{\dagger}_1\hat{a}_1 + \sigma_{1A}^+ \sigma_{1A}^- + \sigma_{2B}^+ \sigma_{2B}^-)$. The Hamiltonian becomes
\begin{eqnarray}
H'_{AB} =&& \Delta_A \sigma_{1A}^+ \sigma_{1A}^-  + \Delta_B \sigma_{2B}^+ \sigma_{2B}^- -g_{1A} (\hat{a}^{\dagger}_1 \sigma_{1A}^-  + \sigma_{1A}^+\hat{a}_1)   - g_{2B} (\hat{a}^{\dagger}_1 \sigma_{2B}^- + \sigma_{2B}^+ \hat{a}_1).
\end{eqnarray}
where $\Delta_A = \omega_{1A}-\omega_{a_1}$ and $\Delta_B = \omega_{2B}-\omega_{a_1}$. We now make a unitary transformation with
\begin{equation}
U = \exp[M] = \exp\Big[\frac{g_{1A}}{\Delta_{A}}(\hat{a}^{\dagger}_1\sigma_{1A}^{-} - \sigma_{1A}^+\hat{a}_1 ) +  \frac{g_{2B}}{\Delta_{B}}(\hat{a}^{\dagger}_1\sigma_{2B}^{-} -  \sigma_{2B}^+\hat{a}_1)\Big].
\end{equation}
We obtain
\begin{eqnarray}
\tilde{H}_{AB}= U H'_{AB} U^{\dagger} = H'_{AB} + [M, H'_{AB}] + \frac{1}{2!}[M,[M,H'_{AB}]] + \ldots
\end{eqnarray}
When the detunings are large, i.e.,
\begin{equation}
g_{1A}, g_{2B} \ll \Delta_{A}, \Delta_{B},
\end{equation}
it is reasonable to consider the effective Hamiltonian to second order in the coupling coefficients $g_{1A}, g_{2B}$. We can then obtain
\begin{eqnarray}
\tilde{H}_{AB}=&&\Big(\Delta_A + \frac{g_{1A}^2}{\Delta_A}\Big) \sigma_{1A}^+\sigma_{1A}^-  +  \Big(\Delta_B + \frac{g_{2B}^2}{\Delta_B}\Big) \sigma_{2B}^+\sigma_{2B}^- + \frac{g_{1A}g_{2B}}{2}\Big(\frac{1}{\Delta_A}+\frac{1}{\Delta_B}\Big) (\sigma_{1A}^+ \sigma_{2B}^- + \sigma_{2B}^+ \sigma_{1A}^-).
\end{eqnarray}
The terms $g_{1A}^2/\Delta_A$ and $g_{2B}^2/\Delta_B$ are the Lamb shifts for atoms $A$ and $B$, respectively. The last term in the above Hamiltonian is the effective coupling between these two atoms. We call it \emph{cross coupling}, because it couples different kinds of atoms. As shown in Fig.~\ref{figS1}(a), the bright blue dashed lines represent cross couplings. For simplicity, we consider $g_{iA}=g_A$ and $g_{iB}=g_B$. The cross coupling is
\begin{equation}
t_c=\frac{g_{A}g_{B}}{2}\Big(\frac{1}{\Delta_A}+\frac{1}{\Delta_B}\Big).
\end{equation}
The couplings between the same atoms can also be implemented. These couplings are called \emph{parallel couplings} for realizing the couplings between the same kinds of atoms. For example, the effective Hamiltonian for atoms $A_1$ and $A_2$ [as shown in Fig.~\ref{figS2}(b)] is
\begin{eqnarray}
\tilde{H}_{AA}=&&\Big(\Delta_A + \frac{\bar{g}_{1A}^2}{\Delta_A}\Big) \sigma_{1A}^+\sigma_{1A}^-  +  \Big(\Delta_A + \frac{\bar{g}_{2A}^2}{\Delta_A}\Big) \sigma_{2A}^+\sigma_{2A}^- + \frac{\bar{g}_{1A}\bar{g}_{2A}}{\Delta_A} (\sigma_{1A}^+ \sigma_{2A}^- + \sigma_{2A}^+ \sigma_{1A}^-).
\end{eqnarray}
As we consider $\bar{g}_{iA}=\bar{g}_A$, the effective coupling between atoms A becomes
\begin{equation}
t_p = \frac{\bar{g}_{A}^2}{\Delta_A}.
\end{equation}

\subsection{Two-resonator coupled three atoms}
\begin{figure}[t]
\includegraphics[width=11cm]{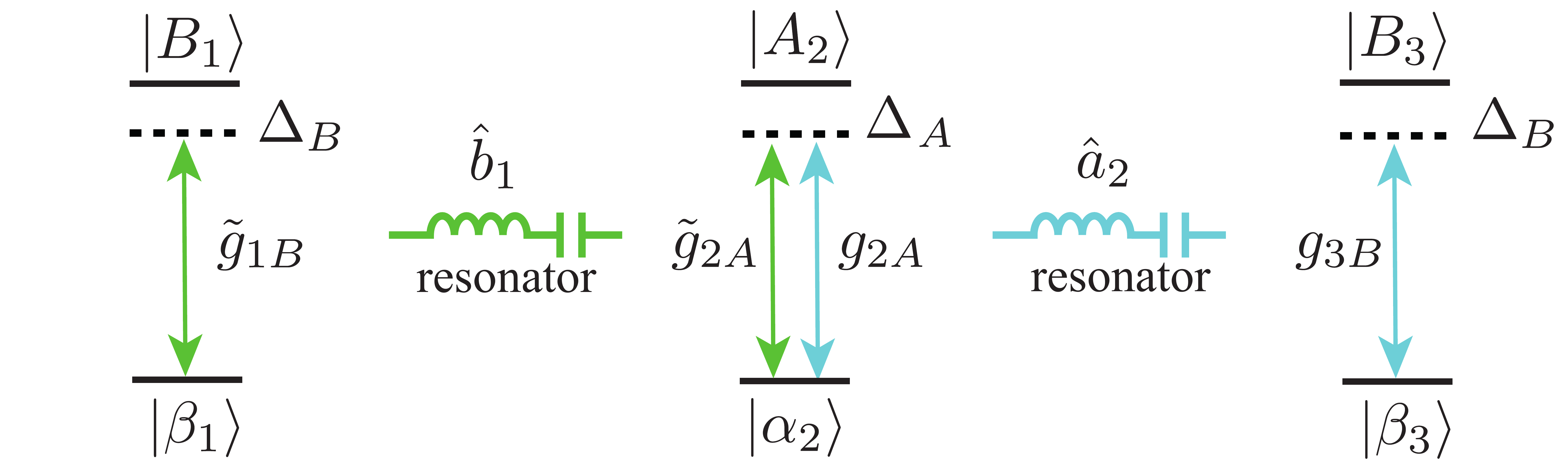}
\caption{Two-resonator-mediated three atoms. }\label{figS3}
\end{figure}

In our system, one atom is coupled to several atoms via different virtual-photons-exchange interactions. We now consider three atoms which are mediated by two \emph{LC} resonators, as shown in Fig.~\ref{figS3}. The corresponding Hamiltonian is
\begin{eqnarray}
H_{BAB} =&& \omega_{b_1} \hat{b}^{\dagger}_1\hat{b}_1 + \omega_{a_2} \hat{a}^{\dagger}_2\hat{a}_2 +  \omega_{B} \sigma_{1B}^+ \sigma_{1B}^-  + \omega_{A} \sigma_{2A}^+ \sigma_{2A}^-  + \omega_{B} \sigma_{3B}^+ \sigma_{3B}^- \nonumber \\
&& -(\tilde{g}_{1B} \hat{b}^{\dagger}_1 \sigma_{1B}^-  +  \tilde{g}_{2A} \hat{b}^{\dagger}_1 \sigma_{2A}^- + \mathrm{H.c.}) - (g_{2A} \hat{a}^{\dagger}_2 \sigma_{2A}^-  +  g_{3B} \hat{a}^{\dagger}_2 \sigma_{3B}^- +\mathrm{H.c.}).
\end{eqnarray}
Here, we assume $\omega_{b_1} = \omega_{a_2} =\omega_r$. Similar to the last section, in the rotating frame, we have
\begin{eqnarray}
H'_{BAB} =&& \Delta_{B} \sigma_{1B}^+ \sigma_{1B}^-  + \Delta_{A} \sigma_{2A}^+ \sigma_{2A}^-  + \Delta_{B} \sigma_{3B}^+ \sigma_{3B}^- \nonumber \\
&& -(\tilde{g}_{1B} \hat{b}^{\dagger}_1 \sigma_{1B}^-  +  \tilde{g}_{2A} \hat{b}^{\dagger}_1 \sigma_{2A}^- + \mathrm{H.c.}) - (g_{2A} \hat{a}^{\dagger}_2 \sigma_{2A}^-  +  g_{3B} \hat{a}^{\dagger}_2 \sigma_{3B}^- + \mathrm{H.c.}),
\end{eqnarray}
where $\Delta_{B}=\omega_{B}-\omega_r$ and $\Delta_{A}=\omega_{A}-\omega_r$. For simplicity, the above Hamiltonian can be expressed as
\begin{equation}
H'_{BAB} = H_1 +H_2 +H_3 + H_{12} + H_{23},
\end{equation}
where $H_{1,2,3}$ are the Hamiltonians for individual atoms. Here, $H_{12}$ and $H_{23}$ are interactions mediated by \emph{LC} resonators. We make a unitary transformation $\tilde{U} = \exp[\tilde{M}]$, with $\tilde{M} =  M_1 + M_2$. Here, $M_1$ and $M_2$ are given by
\begin{eqnarray}
M_1 =  \frac{\tilde{g}_{1B}}{\Delta_{B}}(\hat{b}_1^{\dagger}\sigma_{1B}^{-} - \sigma_{1B}^+ \hat{b}_1) +  \frac{\tilde{g}_{2A}}{\Delta_{A}}(\hat{b}^{\dagger}_1\sigma_{2A}^{-} -  \sigma_{2A}^+\hat{b}_1), \nonumber \\
M_2 = \frac{g_{2A}}{\Delta_{A}}(\hat{a}_2^{\dagger}\sigma_{2A}^{-} - \sigma_{2A}^+\hat{a}_2) + \frac{g_{3B}}{\Delta_{B}}(\hat{a}_2^{\dagger}\sigma_{3B}^{-} - \sigma_{3B}^+\hat{a}_2).
\end{eqnarray}
Then,
\begin{equation}
\tilde{U} H'_{BAB} \tilde{U}^{\dagger} = H'_{BAB} + [\tilde{M},H'_{BAB}] + \frac{1}{2!} [\tilde{M},[\tilde{M},H'_{BAB}]] + \ldots,
\end{equation}
with $[\tilde{M},H'_{BAB}] = [M_1,H'_{BAB}] + [M_2,H'_{BAB}]$. We first consider the term $[M_1,H'_{BAB}]$,
\begin{eqnarray}
[M_1, H'_{BAB}] = [M_1,  H_1 + H_2 + H_{12}] + [M_1, H_{23}].
\end{eqnarray}
We now look at the second term on the right-hand side of the above expression,
\begin{eqnarray}
[M_1, H_{23}] &=& \Big[\frac{\tilde{g}_{2A}}{\Delta_{A}}(\hat{b}^{\dagger}_1\sigma_{2A}^{-} -  \sigma_{2A}^+\hat{b}_1),  - g_{2A}( \hat{a}^{\dagger}_2 \sigma_{2A}^-  + \sigma_{2A}^+ \hat{a}_2)\Big] \nonumber \\
&=&-\frac{\tilde{g}_{2A} g_{2A}}{\Delta_{A}} (\hat{b}^{\dagger}_1 \hat{a}_2 + \hat{a}^{\dagger}_2 \hat{b}_1)(|\alpha_2\rangle \langle \alpha_2|-|A_2\rangle\langle A_2|).
\end{eqnarray}
In our system, the couplers are set to be vacuum states. The real photon exchange can be ignored. Therefore, $[M_1, H_{23}]=0$. So,
\begin{eqnarray}
[\tilde{M},H'_{BAB}] = [M_1, H_1 + H_2 + H_{12}]  + [M_2, H_2 + H_3 + H_{23}].
\end{eqnarray}
Similarly,
\begin{eqnarray}
[\tilde{M},[\tilde{M},H'_{BAB}]] =&& [M_1, [M_1, H_1 +H_2 +H_{12}]] + [M_2, [M_2, H_2 +H_3 +H_{23}]] \nonumber \\
& +& [M_2, [M_1, H_1 +H_2 +H_{12}]] + [M_1, [M_2, H_2 +H_3 +H_{23}]].
\end{eqnarray}
To second order in $g$, we have
\begin{eqnarray}
[M_2, [M_1, H_1 +H_2 +H_{12}]] = 0,
\end{eqnarray}
and
\begin{equation}
[M_1, [M_2, H_2 +H_3 +H_{23}]]=0.
\end{equation}
Hence,
\begin{eqnarray}
[\tilde{M},[\tilde{M},H'_{BAB}]] = [M_1, [M_1, H_1 +H_2 +H_{12}]] + [M_2, [M_2, H_2 +H_3 +H_{23}]].
\end{eqnarray}
Therefore,
\begin{eqnarray}
\tilde{H}_{BAB}&=&\tilde{U} H'_{BAB} \tilde{U}^{\dagger} \nonumber \\
&=& \Big(\Delta_{B}+\frac{\tilde{g}_{1B}^2}{\Delta_B}\Big) \sigma_{1B}^+ \sigma_{1B}^-  + \Big(\Delta_{A} + \frac{\tilde{g}_{2A}^2+g_{2A}^2}{\Delta_A}\Big) \sigma_{2A}^+ \sigma_{2A}^-  + \Big(\Delta_{B}+\frac{g_{3B}^2}{\Delta_B}\Big) \sigma_{3B}^+ \sigma_{3B}^- \nonumber \\
&&+  \frac{\tilde{g}_{1B}\tilde{g}_{2A}}{2}\Big(\frac{1}{\Delta_A}+\frac{1}{\Delta_B}\Big)(\sigma_{1B}^+ \sigma_{2A}^- + \sigma_{2A}^+ \sigma_{1B}^-) +\frac{g_{2A}g_{3B}}{2}\Big(\frac{1}{\Delta_A}+\frac{1}{\Delta_B}\Big)(\sigma_{2A}^+ \sigma_{3B}^- + \sigma_{3B}^+\sigma_{2A}^-). \label{HBAB}
\end{eqnarray}
This effective Hamiltonian shows that the chain-like coupling scheme, as shown in Fig.~\ref{figS3}, does not lead to long-range couplings between atoms. By assuming $\tilde{g}_{1B}=g_{3B}=g_B$ and $\tilde{g}_{2A}=-g_{2A}=-g_A$, Eq.~(\ref{HBAB}) can be written as
\begin{eqnarray}
\tilde{H}_{BAB} &=& \Big(\Delta_{B}+\frac{g_{B}^2}{\Delta_B}\Big) \sigma_{1B}^+ \sigma_{1B}^-  + \Big(\Delta_{A} + 2\frac{g_{A}^2}{\Delta_A}\Big) \sigma_{2A}^+ \sigma_{2A}^-  + \Big(\Delta_{B}+\frac{g_B^2}{\Delta_B}\Big) \sigma_{3B}^+ \sigma_{3B}^- \nonumber \\
&&+ (-t_c\sigma_{1B}^+ \sigma_{2A}^- + t_c \sigma_{2A}^+ \sigma_{3B}^- + \mathrm{H.c.}).
\end{eqnarray}

\subsection{Boundary conditions}
Using the periodic boundary conditions, the translational invariance makes the Lamb shifts for the same kinds of atoms to be equal. We denote the energy splitting between atoms $A$ and $B$ to be $2\delta$. Then, the effective Hamiltonian becomes
\begin{eqnarray}
\tilde{H}= \sum_{i=1}^N \Big[\frac{\delta}{2} (\sigma_{iA}^+ \sigma_{iA}^- - \sigma_{iB}^+ \sigma_{iB}^-) +  t_p (\sigma_{iA}^+ \sigma_{i+1A}^- - \sigma_{iB}^+ \sigma_{i+1B}^- )  - t_c (\sigma_{iA}^+ \sigma_{i+1B}^- - \sigma_{iB}^+ \sigma_{i+1A}^- ) \Big] + \mathrm{H.c.},  \label{effH_SM}
\end{eqnarray}
with $\sigma_{N+1\mu}^{\pm} = \sigma_{1\mu}^{\pm}$ ($\mu=A,B$). In Eq.~(\ref{effH_SM}), the effective couplings have been simplified. The topological property is analysed in the main text. Using open boundary conditions, the atoms of unit cells at the boundaries have different Lamb shifts compared to atoms in other unit cells. However, we can couple vacuum resonators or cavities to these boundary atoms to generate additional Lamb shifts, such that all the atoms of the same kind have the same energy.

\begin{figure}[t]
\includegraphics[width=14cm]{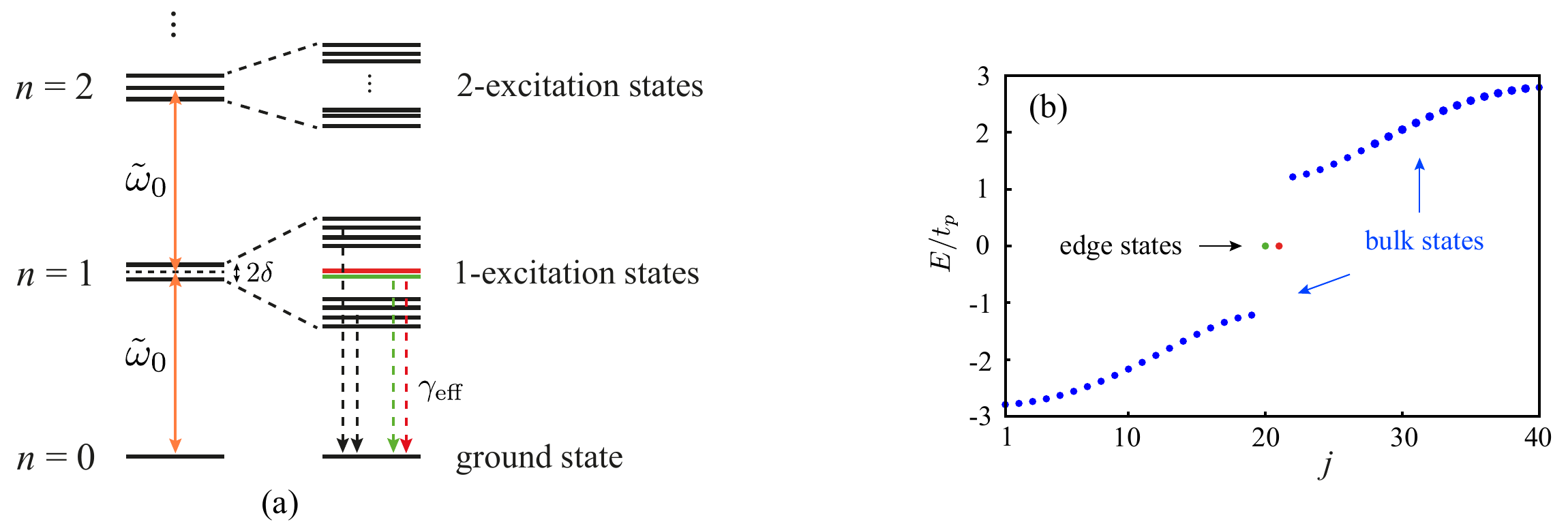}
\caption{Topological superatom. (a) Energy levels for subspaces with different excitations. The energy levels on the left side are produced when $t_c$ and $t_p$ are zero. Here $\tilde{\omega}_0$ represents the middle frequency of atoms $A$ and $B$. On the right side, finite $t_c$ and $t_p$ split the energy degeneracies in single-excitation and multiple-excitation subspaces. Due to $t_c, t_p \ll \tilde{\omega}_0$, the single-excitation subspace is separated from the two-excitation subspace. (b) Large gaps between edge states and bulk states in the single-excitation subspace of the atom array. Here $j$ labels the eigenstates of the atom array in the single-excitation subspace and the number of unit cells is $N=20$.}\label{figS4}
\end{figure}

\section{Topological superatom}
The single-excitation subspace is well-separated from  multiple-excitation subspaces, as shown in Fig.~\ref{figS4}(a). In our model with superconducting quantum circuits, $t_c$ and $t_p$ are tens of MHz, $\tilde{\omega}_0$ is several GHz. We denote $|\mathcal{A}_i\rangle= \sigma_{iA}^+ |G\rangle$ and $|\mathcal{B}_i\rangle= \sigma_{iB}^+ |G\rangle$ with $|G\rangle=|\alpha_1\beta_1\alpha_2\beta_2\cdots\rangle$ being the ground state of the atom array. Then we have
\begin{equation}
\langle \mathcal{A}_i|\sigma_{iA}^+ \sigma_{i+1A}^-|\mathcal{A}_{i+1}\rangle=\langle G|G\rangle =1,
\end{equation}
and similarly $\langle \mathcal{B}_{i}|\sigma_{iB}^+ \sigma_{i+1B}^-|\mathcal{B}_{i+1}\rangle=1$, $\langle \mathcal{A}_{i}|\sigma_{iA}^+ \sigma_{i+1B}^-|\mathcal{B}_{i+1}\rangle=1$, $\langle \mathcal{B}_{i}|\sigma_{iB}^+ \sigma_{i+1A}^-|\mathcal{A}_{i+1}\rangle=1$. Therefore, the Hamiltonian Eq.(3) in the main text can be written in the single-excitation subspace $\{|\mathcal{A}_i\rangle, |\mathcal{B}_i\rangle\}$ as
\begin{equation}
\bar{H}= \sum_{i=1}^N \Big[\frac{\delta}{2} (|\mathcal{A}_i\rangle\langle \mathcal{A}_i| - |\mathcal{B}_i\rangle\langle \mathcal{B}_i|) +  t_p (|\mathcal{A}_i\rangle\langle \mathcal{A}_{i+1}| - |\mathcal{B}_i\rangle\langle \mathcal{B}_{i+1}| )  - t_c (|\mathcal{A}_i\rangle\langle \mathcal{B}_{i+1}| - |\mathcal{B}_i\rangle\langle \mathcal{A}_{i+1}| ) \Big] + \mathrm{H.c.}.
\end{equation}
In crystal momentum space, the Hamiltonian becomes $\bar{H}(k) = \sum_{k} \Psi_k^{\dagger} h(k) \Psi_k$, with $\Psi_k^{\dagger} = (|\mathcal{A}_k \rangle,|\mathcal{B}_k \rangle)$, and
\begin{equation}
h(k) = d_y(k) \sigma_y + d_z(k) \sigma_z,
\end{equation}
where $d_y(k) = 2 t_c \sin k$ and $d_z(k) = \delta + 2 t_p \cos k$.

Edge states are topologically protected quantum many-body states. They are able to encode quantum information and can be used as topological qubits. Recently, the study of  Majorana zero modes has advanced considerably. Topological quantum computation can be potentially implemented with Majorana fermions. There are theoretical proposals suggesting that photon-electron interactions could be used to control Majorana fermions. However, the photon-electron interactions are not easy to control, compared to light-atom interactions. Especially, in some artificial atoms, e.g., superconducting quantum circuits, one can optically manipulate quantum states of atoms with high accuracy.

The 1D atom array studied here has a complex energy spectrum. In its topological phase, as shown in Fig.~\ref{figS4}(b), bulk states exhibit a smooth spectrum with very small gaps among the bulk states. This makes it difficult to address specific quantum many-body states. However, there are large gaps between the two $E=0$ edge states and bulk states. This provides a strong nonlinearity to control the edge states. In quantum systems, the nonlinearity of energy levels is critical for qubits or qutrits, where quantum information can be encoded. Due to the large gaps between edge states and bulk states, the topological superatom with a ground state and two edge states can be used to characterize the atom array. We can exploit the properties of edge states, i.e., topology-protected spin polarization and boundary localization, to implement the interaction between light and the topological superatom. Benefiting from the atom-light couplings, which are studied in many quantum optical systems, the topological superatom could be easily addressed.

\subsection{Edge states in the single-excitation subspace}
The atom array mediated by couplers is shown to have topological structure in crystal momentum space. From the edge-bulk correspondence, edge states can be generated in the topological phase with open boundary conditions. Different from normal many-body states, edge states have peculiar properties that can be employed for topological quantum state engineering. Therefore, we better analyze the wavefunctions of edge states. In the single-excitation subspace, the Hamiltonian can be written as \cite{Konig2008,PhysRevA.81.033622S},
\begin{eqnarray}
\bar{H}=\sum_{i=1} ^{N} \mathcal{M}\Psi_i^{\dagger} \Psi_i + \mathcal{T}^{\dagger} \Psi_{i+1}^{\dagger} \Psi_i  +  \mathcal{T} \Psi_i^{\dagger} \Psi_{i+1} , \label{Hbar}
\end{eqnarray}
with
\begin{equation}
\mathcal{M} = \delta \sigma_z, \quad\quad
\mathcal{T} = t_p \sigma_z +i t_c \sigma_y. \nonumber
\end{equation}
We now make an ansatz for the edge state $\psi=\sum_n \lambda^n \phi$, where $\phi$ is a 2 component spinor. Therefore,
\begin{equation}
\bar{H}  \psi = E \psi.
\end{equation}

\begin{figure}[b]
\includegraphics[width=7cm]{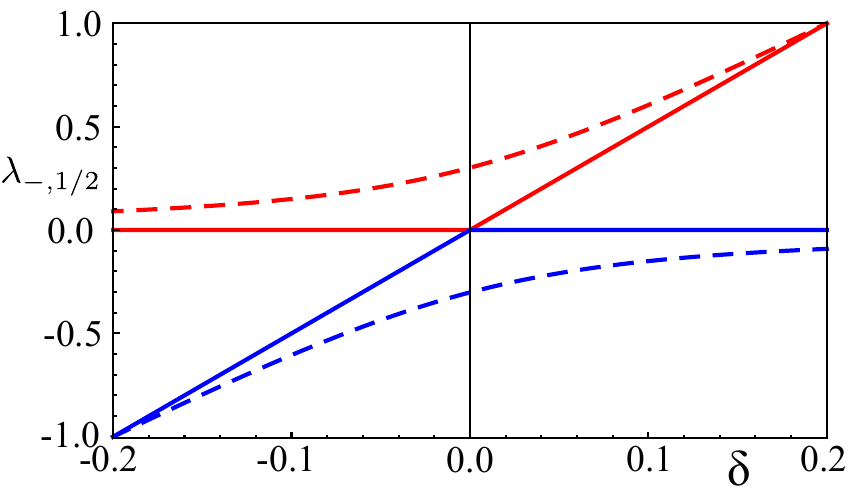}
\caption{The eigenvalue $\lambda$ of $\phi_-$ (see Eq.~(\ref{lambdam})) versus $\delta$, with $t_p=-0.1$. The red-solid and red-dashed curves represent $\lambda_{-,1}$ with different values of $t_c$: $-0.1$ and $-0.12$, respectively. Similarly, the blue-solid and blue-dashed curves show $\lambda_{-,2}$ for $t_c=-0.1$ and $t_c=-0.12$, respectively. The effective energy difference between the two excited states $|A_i\rangle$ and $|B_i\rangle$ of the \emph{i}th atom is $2\delta$.}\label{figS5}
\end{figure}

From the above equation, we can have
\begin{equation}
(\mathcal{M} + \lambda \mathcal{T}^{\dagger} + \lambda^{-1} \mathcal{T} ) \phi =E \phi. \label{eq1}
\end{equation}
This can be written as
\begin{equation}
[\delta \sigma_z + \lambda (t_p \sigma_z - i t_c \sigma_y)+ \lambda^{-1} (t_p \sigma_z + i t_c \sigma_y)] \phi = E \phi.
\end{equation}
The edge states are solutions with $E=0$, i.e.,
\begin{equation}
[\delta \sigma_z + \lambda (t_p \sigma_z - i t_c \sigma_y)+ \lambda^{-1} (t_p \sigma_z + i t_c \sigma_y)] \phi =0.
\end{equation}
Multiplying $\sigma_z$ from the left-hand side, one obtains
\begin{equation}
[\delta  + \lambda (t_p  - t_c \sigma_x) + \lambda^{-1} (t_p+ t_c \sigma_x)] \phi =0. \label{edge_H}
\end{equation}
We can obtain the eigenstates $\phi_{\pm}$ via
\begin{equation}
\sigma_x \phi_{\pm} = \pm \phi_{\pm}. \label{phipm}
\end{equation}
From Eq.~(\ref{edge_H}), we can have $\delta  + \lambda (t_p  - t_c) - \lambda^{-1} (-t_p - t_c) =0$, which is a quadratic equation for $\lambda$. It can be solved with solutions,
\begin{eqnarray}
\lambda_{+,1} =  \frac{\delta + \sqrt{\delta^2 + 4(t_c^2 - t_p^2)} }{2(t_c - t_p)}, \quad\quad
\lambda_{+,2} =  \frac{\delta - \sqrt{\delta^2 + 4(t_c^2 - t_p^2)} }{2(t_c - t_p)}, \label{lambdap}
\end{eqnarray}
for $\phi_+$, and
\begin{eqnarray}
\lambda_{-,1} =  \frac{\delta + \sqrt{\delta^2 + 4(t_c^2 - t_p^2)} }{2(-t_c - t_p)}, \quad\quad
\lambda_{-,2} =  \frac{\delta - \sqrt{\delta^2 + 4(t_c^2 - t_p^2)} }{2(-t_c - t_p)}. \label{lambdam}
\end{eqnarray}
for $\phi_-$. The values of $\lambda_{\pm,1/2}$ determine the wavefunctions of the edges states. From Eq.~(\ref{lambdap}) and Eq.~(\ref{lambdam}), we can find that $1/\lambda_{+,1}=\lambda_{-,2}$ and $1/\lambda_{+,2}=\lambda_{-,1}$. So, there are two cases that lead to different edge states in the system.

Case (1): If $|\lambda_{+,1}| <1$ and $|\lambda_{+,2}| <1$, the edge state of the left boundary is polarized along $\phi_{+}$. The component of the wavefunction in the $i$th unit cell is
\begin{equation}
\psi_L(i) = \left[c_1 (\lambda_{+,1})^i + c_2 (\lambda_{+,2})^i\right] \phi^{(i)}_{+}.
\end{equation}
The open boundary condition requires the amplitude of $\psi_L(0)$ to be zero, which gives $c_1 = -c_2$. Therefore, the left edge state is
\begin{equation}
\psi_L =\frac{1}{\sqrt{\mathcal{N}_L^+}}\sum_i \left[(\lambda_{+,1})^i - (\lambda_{+,2})^i\right] \phi^{(i)}_{+}, \label{wavefunction1}
\end{equation}
where $\mathcal{N}_L^+$ is the normalization factor. Similarly, the right edge state with open boundary condition is
\begin{equation}
\psi_R =\frac{1}{\sqrt{\mathcal{N}_R^-}} \left[ (\lambda_{+,1})^{N+1-i} - (\lambda_{+,2})^{N+1-i}\right] \phi^{(i)}_{-}.
\end{equation}
Case (2): If $|\lambda_{+,1}| >1$ and $|\lambda_{+,2}| >1$, the edge state of the left boundary is polarized along $\phi_{-}$, because of $1/\lambda_{+,1} = \lambda_{-,2}$ and $1/\lambda_{+,2} = \lambda_{-,1}$. The wavefunctions for the left and right edge states are
\begin{equation}
\psi_L = \frac{1}{\sqrt{\mathcal{N}_L^-}}\sum_i\left[ (\lambda_{-,1})^i - (\lambda_{-,2})^i\right] \phi^{(i)}_{-},
\end{equation}
and
\begin{equation}
\psi_R = \frac{1}{\sqrt{\mathcal{N}_R^+}}\sum_i\left[ (\lambda_{-,1})^{N+1-i} - (\lambda_{-,2})^{N+1-i}\right] \phi^{(i)}_{+}. \label{wavefunction2}
\end{equation}
The values of $\lambda_{\pm,1/2}$, which are determined by the system parameters, affect the form of the edge states. In Fig.~\ref{figS5}, we show $\lambda_{-,1/2}$ for two cases, i.e., $|t_c|=|t_p|$ and $|t_c| \neq |t_p|$ in the topological phase ($-2|t_p|<\delta<2|t_p|$). As $|t_c|=|t_p|$, only one parameter, $\lambda_{-,1}$ or $\lambda_{-,2}$ is nonzero. However, in the case of $|t_c| \neq |t_p|$, both $\lambda_{-,1}$ and $\lambda_{-,2}$ are nonzero. This two different forms of edge states have distinctive features in the finite-size effects of the edge states, as we show in the main text.

\begin{figure}[b]
\includegraphics[width=14cm]{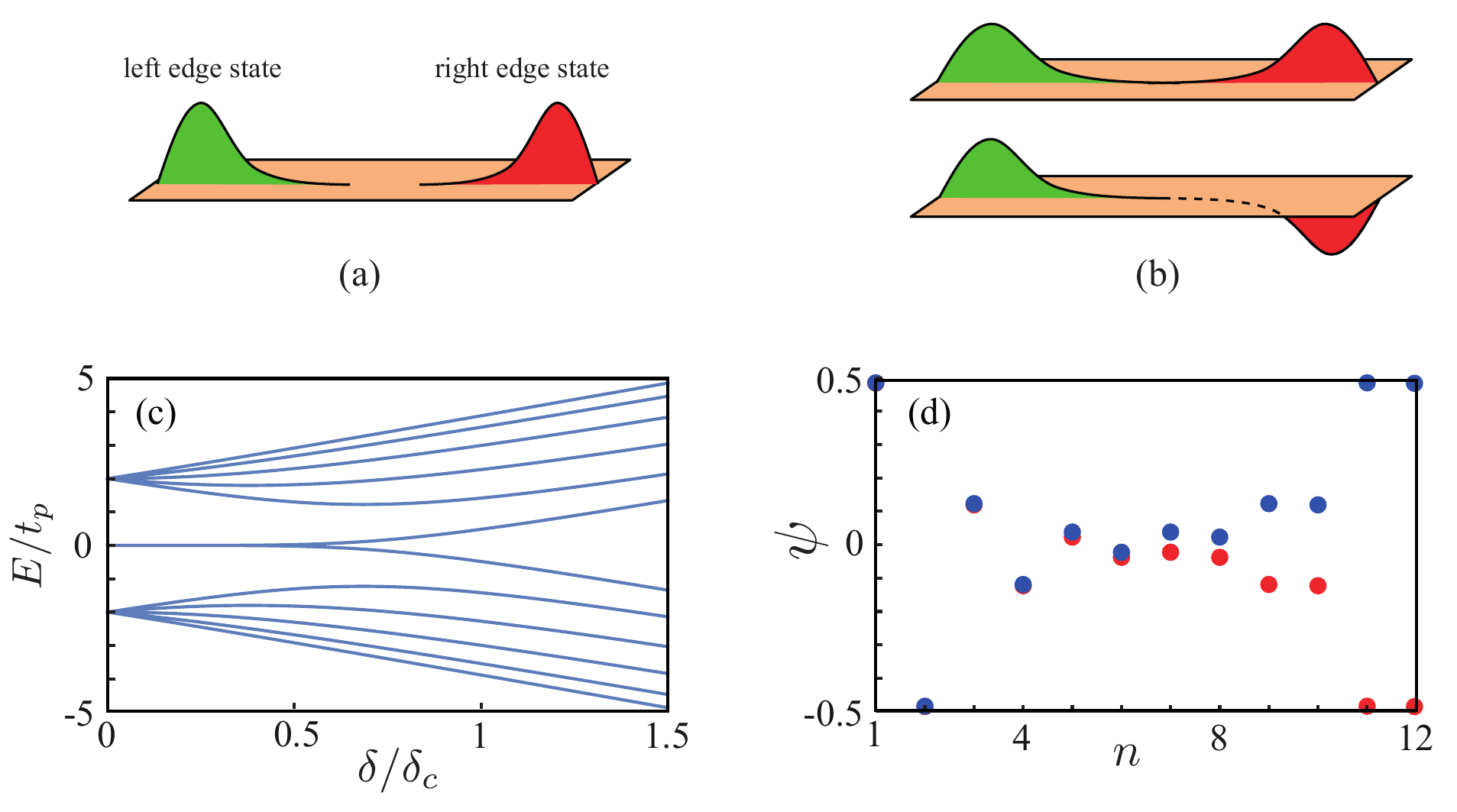}
\caption{(a) The left and right edge states are well separated in large arrays. (b) Edge states are hybridized due to finite size of the atom array. (c) Energy spectrum of the atom array with $N=6$ unit cells. (d) Wavefunction of hybridized edge states for $\delta/\delta_c=0.25$. Here $n$ labels the positions of atoms in the array, i.e., odd (even) number of $n$ corresponds to $|\mathcal{A}_{\frac{n+1}{2}}\rangle$ ($|\mathcal{B}_{\frac{n}{2}}\rangle$).}\label{figS6}
\end{figure}
The edge states shown above, i.e., Eqs.~(\ref{wavefunction1})-(\ref{wavefunction2}), describe long lattices, as shown in Fig.~\ref{figS6}(a). For short lattices, the edge states are not separated, but hybridize with each other, as shown in Fig.~\ref{figS6}(b). The hybridized edge states may have interesting observable effects. The hybridization leads to splitting between edge states (see Fig.~\ref{figS6}(c)). As shown in Fig.~\ref{figS6}(d), the hybridized edge states can be written as $\psi_{\pm}=\frac{1}{\sqrt{2}}(\psi_L \pm \psi_R)$, where $\psi_L$ and $\psi_R$ are the left and right localized edge states, respectively. We consider left edge state to be the initial state, i.e., $\psi_0=\frac{1}{\sqrt{2}}(\psi_+ + \psi_-)$. The evolution of the system is
\begin{eqnarray}
\psi(t)&=&\frac{1}{\sqrt{2}}e^{-i \tilde{H} t/\hbar} (\psi_+ + \psi_-) \nonumber \\
&=&\frac{1}{\sqrt{2}}e^{-i \tilde{\omega}_0 t} (e^{-i \Delta_s t} \psi_+\ + e^{i \Delta_s t} \psi_- ) \nonumber \\
&=& e^{-i \tilde{\omega}_0 t}\Big[\frac{1}{\sqrt{2}}\cos(\Delta_s t)(\psi_+ + \psi_-) -i \frac{1}{\sqrt{2}}\sin(\Delta_s t)(\psi_+ - \psi_-)\Big] \nonumber \\
&=& e^{-i \tilde{\omega}_0 t}[\cos(\Delta_s t)\psi_L -i \sin(\Delta_s t) \psi_R],
\end{eqnarray}
where $\tilde{\omega}_0$ is the middle frequency of two edge states, and $\Delta_s$ represents the splitting between them. When $\Delta_s \approx 0$, the excitation localizes to the left edge of the atom array. Otherwise, the excitation oscillates between the left-edge and right-edge atoms. This oscillatory behavior shows the interaction between edge states. Since the atoms at left and right edges are respectively subradiant and superradiant, the excitation mainly relaxes from the right edge. And from the revival of the excitation, we can estimate the decay rate of the superradiant (right) edge state. For the case $\delta/\delta_c=0.25$, the value of $\Delta_s$ becomes $4.6\times 10^{-5}\kappa$. Because of $\Delta_s \ll \gamma$, the edge states are localized during the lifetime of single atoms. However, when $\Delta_s$ becomes large ($\Delta_s \approx \gamma$), the population dynamics of the subradiant edge state can be used to measure the coupling strength of the edge states.

\subsection{Driving a topological superatom in a cavity}
In our one-dimensional topological array with V-shaped effective three-level atoms, the edge states are produced in the excited state. Moreover, thanks to symmetry protection, there are many features unique to edge states, i.e., spin polarization, boundary localization, and large energy gaps to bulk states. These properties make it feasible to optically manipulate edge states. The coupling between ground and excited edge states can be realized by choosing appropriate cavity-atom coupling parameters, such that the edge states are efficiently populated.  For example, in superconducting quantum circuits, the couplings between artificial atoms and cavity can be controlled. Therefore, the topological superatom can be controlled. Here, we consider the low-excitation limit, i.e., $\langle \sigma_{i\alpha}^+\sigma_{i\alpha}^- \rangle \approx 0$, with $\alpha=A,B$. The master equation of the cavity-driving atom array is
\begin{equation}
\dot{\rho} = i [\rho,H_{\mathrm{tot}}] + \mathcal{L}_a[\rho] + \mathcal{L}_c[\rho],
\end{equation}
with total Hamiltonian $H_{\mathrm{tot}} =\tilde{H}+ H_c + H_I$. Here, $\tilde{H}$ represents the coupler-mediated atom array, $H_c$ is the Hamiltonian of the cavity, and $H_I$ is the cavity-atom interaction. The dissipation terms for the atom array and cavity are
\begin{equation}
\mathcal{L}_a[\rho] = \sum_{i,\mu,\nu} \gamma_{\mu\nu} (2\sigma_{i\mu}^- \rho \sigma_{i\nu}^+ - \sigma_{i\mu}^+ \sigma_{i\nu}^- \rho - \rho \sigma_{i\mu}^+ \sigma_{i\nu}^-),
\end{equation}
and
\begin{equation}
\mathcal{L}_c[\rho] = \kappa(2 \hat{f} \rho \hat{f}^{\dagger}-\hat{f}^{\dagger} \hat{f} \rho - \rho \hat{f}^{\dagger} \hat{f}),
\end{equation}
respectively. From the master equation, we obtain the equations
\begin{eqnarray}
\big\langle \frac{d}{dt} \hat{f}\big\rangle &=& -(\kappa + i \Delta_c) \langle \hat{f} \rangle -i \bm{\mathrm{\Xi}}^{\mathrm{T}} \langle \bm{\sigma} \rangle + \eta,  \\
\big\langle \frac{d}{dt}\bm{\sigma} \big\rangle &=& -i (\bm{\Delta} + \bm{D} -i \bm{\Gamma}) \langle \bm{\sigma} \rangle -i \bm{\mathrm{\Xi}} \langle \hat{f} \rangle,
\end{eqnarray}
with $\bm{\mathrm{\Xi}}=(\xi_{1A},\xi_{1B},\xi_{2A},\xi_{2B}, \cdots)$, $\langle \bm{\sigma} \rangle=(\langle \sigma_{1A}^- \rangle, \langle \sigma_{1B}^- \rangle, \langle \sigma_{2A}^- \rangle, \langle \sigma_{2B}^- \rangle, \cdots)^{\mathrm{T}}$, $\bm{\Delta}=\mathrm{Diag}(\delta,-\delta,\delta,-\delta,\cdots)$,
\begin{eqnarray}
\bm{\Gamma} = \left(
                \begin{array}{ccccc}
                  \gamma_{1A} & \gamma_{1AB} & 0 & 0 & 0 \\
                  \gamma_{1AB} & \gamma_{1B} & 0 & 0 & 0 \\
                  0 & 0 & \gamma_{2A} & \gamma_{2AB} & 0 \\
                  0 & 0 & \gamma_{2AB} & \gamma_{2B} & 0 \\
                  0 & 0 & 0 & 0 & \ddots \\
                \end{array}
              \right),
\end{eqnarray}
and
\begin{eqnarray}
\bm{D} = \left(
           \begin{array}{cccc}
             0 & R & 0 & 0 \\
             R^T & 0 & R & 0 \\
             0 & R^T & 0 & \ddots \\
             0 & 0 & \ddots & \ddots \\
           \end{array}
         \right),
\end{eqnarray}
where
\begin{eqnarray}
R = \left(
        \begin{array}{cc}
          t_p & -t_c \\
          t_c & -t_p \\
        \end{array}
      \right). \nonumber
\end{eqnarray}
Here, $\xi_{i\alpha}$ are the coupling coefficients between the atoms and cavity. The steady cavity field can be solved by assuming $\langle \frac{d}{dt}\hat{f} \rangle=0$ and $\langle \frac{d}{dt}\bm{\sigma} \rangle=0$. Then, we can obtain the transmission
\begin{equation}
T=|t|^2=\Big|\frac{\kappa}{\kappa + i \Delta_c -i \chi}\Big|^2, \label{transmission}
\end{equation}
with
\begin{equation}
t= \kappa \langle \hat{f} \rangle/\eta
\end{equation}
and susceptibility
\begin{equation}
\chi=\bm{\Xi}^\intercal (\Delta + \bm{D} -i \bm{\Gamma})^{-1} \bm{\Xi}.
\end{equation}
\begin{figure}[b]
\includegraphics[width=14cm]{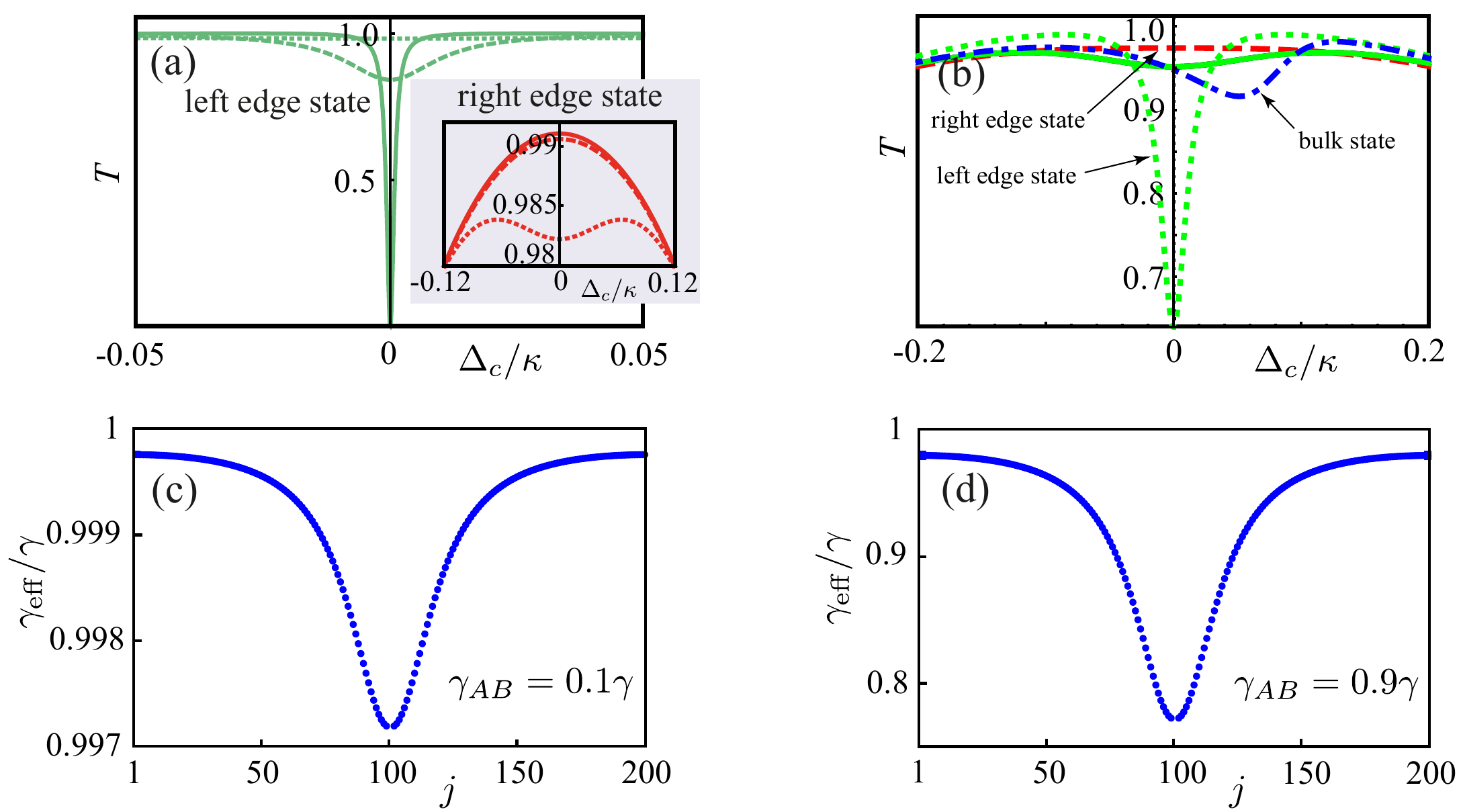}
\caption{(a) Transmission spectra of the left and right (the inset) edge states. Here, $\Delta_c = \omega_{\mathrm{cavity}}-\omega_{\mathrm{drive}}$ and $\delta/\delta_c=0.6$. Dotted, dashed, and solid lines correspond to $\gamma_{AB}/\gamma=0, 0.9, 1$, respectively. (b) Transmission  spectra of the two edge states and one bulk state. The red-dashed(-solid), green-dotted(-solid), and blue-dot-dashed(-solid) curves are the transmissions for edge states and bulk state, with $\gamma_{AB}/\gamma=0.9$ ($\gamma_{AB}/\gamma=0$), $\delta/\delta_c=0.65$. (c,d) Effective decays of bulk states for $\gamma_{AB}=0.1\gamma$ and  $\gamma_{AB}=0.9\gamma$, respectively, with $\delta/\delta_c=0.6$.  The horizontal axis $j$ represents bulk states from lowest energy to highest energy states (edge states with $j=100$ and $j=101$ are not shown). Other parameters for these figures are $N=100, t_c=t_p,\kappa=10\gamma$.}\label{figS7}
\end{figure}
When a quantum many-body state is driven by the cavity field, one can probe its optical response via its photon transmission. The susceptibility captures the central property of the cavity-driving many-body system. From the susceptibility, we can obtain the effective decay rate of the superatom,
\begin{equation}
\gamma_{\mathrm{eff}}=-\mathrm{Im}\Big[\frac{\bm{\Xi}^\intercal \bm{\Xi}}{\chi}\Big].
\end{equation}
In particular, the edge states in the single-excitation subspace have zero energy, which makes
$\mathrm{Re}[\chi]$ vanishing. When the edge state is resonantly driven, the transmission can be expressed by the effective decay
\begin{equation}
T_{\mathrm{res}}=\frac{\kappa^2}{(\kappa+\mathrm{Im}[\chi])^2}.
\end{equation}
The invariance of $\mathrm{Im}[\chi]$ for edge states indicates the topologically protected quantum coherence. As shown in Fig.~3(b), the bulk states in the non-topological phase also have constant $\mathrm{Im}[\chi]$ when $\delta$ is large. This represents that the decay rates of bulk states have an upper bound $\gamma$. In the main text, we consider that the cavity has low decay rate $\kappa$, i.e., $\kappa=0.1\gamma$. The cavity decay $\kappa$ plays important role in the transmission of edge states. In Fig.~\ref{figS7}(a), we consider a large cavity decay. The left edge state has clear signal as $\gamma_{AB}$ increases. However, the transmission for right edge state is not changed so much. In Fig.~\ref{figS7}(b), the transmission spectra for two edge states and one bulk state are compared. When $\gamma_{AB}$ is zero, the transmissions for edge and bulk states are the same. When $\gamma_{AB}$ is nonzero, the spectrum is found to be asymmetric for bulk state, but symmetric for edge states.

As shown in the main text, the effective decay rates for bulk states and edge states are equal to $\gamma$ for $\gamma_{AB}=0$. However, nonzero correlated decay $\gamma_{AB}$ makes the bulk states to be subradiant. In Figs.~\ref{figS7}(c) and \ref{figS7}(d), we show the effective decays for bulk states with different values of $\gamma_{AB}$. The $x$ axis denotes the index of the bulk states, from lowest energy to the largest (the edge states for $n=N,N+1$ in the middle are not shown). It can be seen that the effective decays for bulk states are symmetric. Moreover, the bulk states closer to edge states are more subradiant. For large correlated decay $\gamma_{AB}$, the bulk states have very different coherence properties compared with edge states; the symmetric edge state is superradiant, and the anti-symmetric edge state is very subradiant. The coherence differences between edge states and bulk states lead to distinctive collective behavior of edge atoms and bulk atoms.

\section{Effects of symmetry breaking and disorders}
\begin{figure}[b]
\includegraphics[width=14cm]{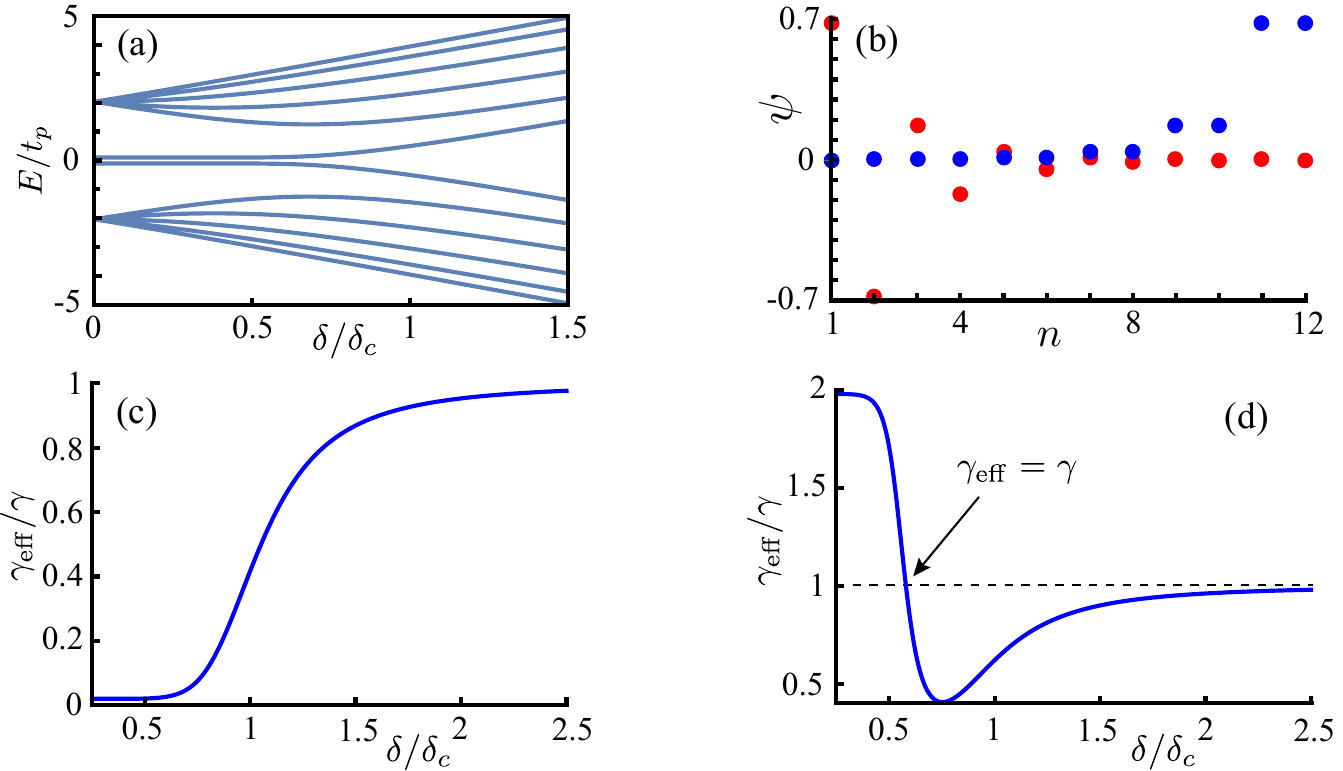}
\caption{Effects of interactions between atoms in the same unit cells. (a) The energy degeneracy for edge states is shifted. (b) The hybridized edge state (see Fig.~\ref{figS6}(d)) become separated (with $\delta=0.25\delta_c$). (c) Quantum coherence of the left edge state. (d) Quantum coherence of the right edge state. The parameters we considered here are $t_c=t_p$, $g_{AB}=0.1\gamma$ and $N=6$.}\label{figbs}
\end{figure}
The waveguides mediate the correlated decays between atoms in unit cells. As shown in Eq.~(\ref{Eq_gAB}) and Fig.~\ref{figS1}(c), when the separation between two atoms along the waveguide coupling them is $2\pi mc/\omega_0$ ($m$ is an integer number), the interaction between these two atoms becomes zero. In experiments, there could be imperfections, such that the separation between two atoms along the waveguide is not exactly $2\pi mc/\omega_0$. If these interactions are homogeneous, i.e., the interactions between atoms in the same unit cells are $g_{AB}$, the Hamiltonian of the system becomes
\begin{eqnarray}
\tilde{H}'&=& \sum_{i=1}^N \delta (\sigma_{iA}^+ \sigma_{iA}^- - \sigma_{iB}^+ \sigma_{iB}^-) + g_{AB} (\sigma_{iA}^+ \sigma_{iB}^- + \sigma_{iB}^+ \sigma_{iA}^-) \nonumber \\
&+&  \sum_{i=1}^{N-1} \Big[t_p (\sigma_{iA}^+ \sigma_{i+1A}^- - \sigma_{iB}^+ \sigma_{i+1B}^- )  - t_c (\sigma_{iA}^+ \sigma_{i+1B}^- - \sigma_{iB}^+ \sigma_{i+1A}^- )   + \mathrm{H.c.} \Big].
\end{eqnarray}
In the crystal momentum space, the Hamiltonian is $\bar{H}'(k) = \sum_{k} \Psi_k^{\dagger} h'(k) \Psi_k$, with
\begin{equation}
h'(k) = g_{AB} \sigma_x + d_y(k) \sigma_y + d_z(k) \sigma_z.
\end{equation}
Apparently, the interactions between atoms in the same unit cells break the chiral symmetry. Accordingly, the energy degeneracy between left- and right-edge states is shifted, as shown in Fig.~\ref{figbs}(a). However, the edge polarizations are preserved, as shown in Fig.~\ref{figbs}(b). Different from Fig.~\ref{figS6}(d), the edge states are not hybridized at $\delta=0.25\delta_c$. The breaking of energy degeneracy for edge states have a nontrivial influence on the topological phase transition. In Figs.~\ref{figbs}(c) and \ref{figbs}(d), we show the effective decays for left and right edge states and their transitions to bulk states. Different from the case with chiral symmetry we discussed in the main text, here there is no interaction between edge states during the topological phase transition. And the topological superradiance-subradiance transition, i.e., $\gamma_{\mathrm{eff}}=\gamma$ as shown in Fig.~\ref{figbs}(d), is produced by the direct edge-bulk transition.

\begin{figure}[b]
\includegraphics[width=14cm]{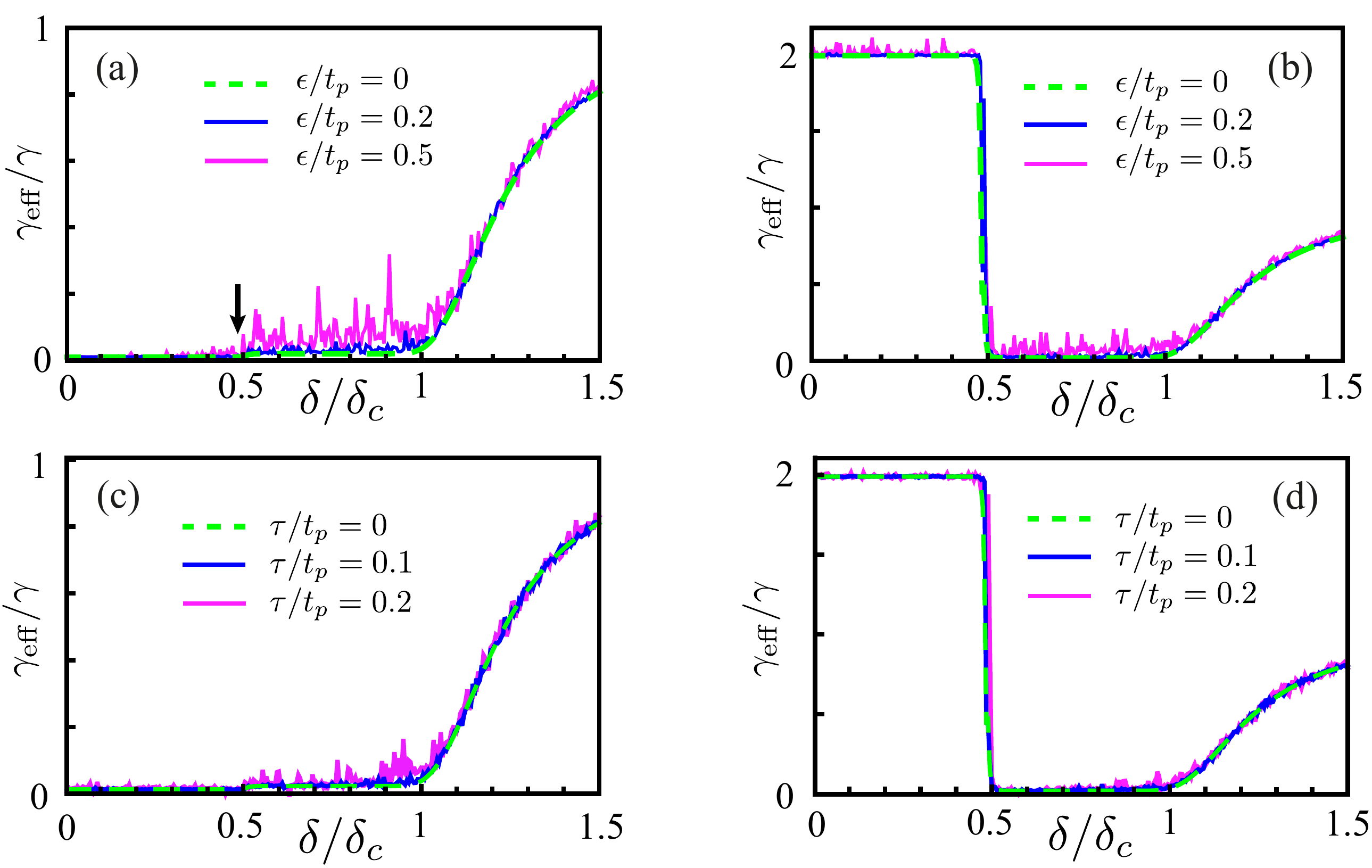}
\caption{Disorders of atomic frequencies for (a) subradiant edge state and (b) superradiant edge state. Disorders of atomic interactions for (c) subradiant edge state and (d) superradiant edge state. The parameters are $t_c=t_p, \gamma=10\kappa, \gamma_{AB}=0.99\gamma, N=50$.}\label{figdisorder}
\end{figure}

In Figs.~\ref{figdisorder}(a) and \ref{figdisorder}(b), we study the effect of disorder of atomic frequencies for the subradiant and superradiant edge states, respectively. The atomic frequencies are $\omega_{i\alpha}+\epsilon_{i\alpha}$ ($\alpha=A,B$), where $\epsilon_{i\alpha}$ are uniformly distributed $\epsilon_{i\alpha} \in [-\epsilon, \epsilon]$. Here, $\epsilon$ represents the strength of the disorder. The unhybridized subradiant edge state ($\delta<\delta_m$) is stable to the noise. However, the hybridized subradiant edge states ($\delta_m<\delta<\delta_c$) and subradiant bulk states in the non-topological phase ($\delta>\delta_c$) are more sensitive to the frequency noise. Similar results are found for the noise of atomic interactions, as shown in Figs.~\ref{figdisorder}(c) and \ref{figdisorder}(d), where the disorder of atomic interactions of strength $\tau$ for the subradiant and superradiant edge states are respectively considered. From Figs.~\ref{figdisorder}(a)-\ref{figdisorder}(d), we find the robustness of quantum coherence to noises for the unhybridized edge states ($\delta<\delta_m$). And the effect of the noises is enhanced for hybridized edge states ($\delta_m<\delta<\delta_c$).


\begin{thebibliography}{99}%
\bibitem{RevModPhys.82.3045} M. Z. Hasan and C. L. Kane, \emph{Colloquium: Topological insulators}, Rev. Mod. Phys. \textbf{82}, 3045 (2010).

\bibitem{RevModPhys.83.1057} X.-L. Qi and S.-C. Zhang, \emph{Topological insulators and superconductors}, Rev. Mod. Phys. \textbf{83}, 1057 (2011).

\bibitem{Bliokh2015Spin} K. Y. Bliokh, F. J. Rodr\'{i}guez-Fortu\~{n}o, F. Nori  and A. V. Zayats, \emph{Spin-orbit interactions of light}, Nat. Photonics \textbf{9}, 796 (2015).

\bibitem{RevModPhys.91.015006} T. Ozawa, H. M. Price, A. Amo, N. Goldman, M. Hafezi, L. Lu, M. Rechtsman, D. Schuster, J. Simon, O. Zilberberg  and I. Carusotto, \emph{Topological Photonics}, Rev. Mod. Phys. \textbf{91}, 015006 (2019).

\bibitem{PhysRevLett.108.220401} L.-J. Lang, X. Cai and S. Chen, \emph{Edge States and Topological Phases in One-Dimensional Optical Superlattices}, Phys. Rev. Lett. \textbf{108}, 220401 (2012).

\bibitem{PhysRevLett.108.255303} N. Goldman, J. Beugnon and F. Gerbier, \emph{Detecting Chiral Edge States in the Hofstadter Optical Lattice}, Phys. Rev. Lett. \textbf{108}, 255303 (2012).

\bibitem{xiaopengli2013} X. Li, E. Zhao and W. V. Liu, \emph{Topological states in a ladder-like optical lattice containing ultracold atoms in higher orbital bands}, Nat. Commun. \textbf{4}, 1523 (2013).

\bibitem{Kitaev2001} A. Y. Kitaev, \emph{Unpaired Majorana fermions in quantum wires}, Phys.-Usp. \textbf{44}, 131 (2001).

\bibitem{PhysRevB.81.014505} J. Q. You, X.-F. Shi, X. Hu and F. Nori, \emph{Quantum emulation of a spin system with topologically protected ground states using superconducting quantum circuits}, Phys. Rev. B \textbf{81}, 014505 (2010).

\bibitem{Alicea2011} J. Alicea, Y. Oreg, G. Refael, F. von Oppen and M. P. A. Fisher, \emph{Non-Abelian statistics and topological quantum information processing in 1D wire networks}, Nat. Phys. \textbf{7}, 412 (2011).

\bibitem{PhysRevLett.110.076401} X.-J. Liu, Z.-X. Liu and M. Cheng, \emph{Manipulating Topological Edge Spins in a One-Dimensional Optical Lattice}, Phys. Rev. Lett. \textbf{110}, 076401 (2013).

\bibitem{2014You} J. Q. You, Z. D. Wang, W. X. Zhang and F. Nori, \emph{Encoding a qubit with Majorana modes in superconducting circuits}, Sci. Rep. \textbf{4}, 5535 (2014).

\bibitem{2011arXiv1110.3788Y} N. Y. Yao, C. R. Laumann, A. V. Gorshkov, H. Weimer, L. Jiang, J. I. Cirac, P. Zoller and M. D. Lukin, \emph{Topologically protected quantum state transfer in a chiral spin liquid}, Nat. Commun. \textbf{4}, 1585 (2013).

\bibitem{dlaska2017robust} C. Dlaska, B. Vermersch and P. Zoller, \emph{Robust quantum state transfer via topologically protected edge channels in dipolar arrays}, Quantum Sci. Technol. \textbf{2}, 015001 (2017).

\bibitem{Bliokh2015Quantum} K. Y. Bliokh, D. Smirnova and F. Nori, \emph{Quantum spin Hall effect of light}, Science \textbf{348}, 1448 (2015).


\bibitem{Bliokh2019Topological} K. Y. Bliokh, D. Leykam, M. Lein and F. Nori, \emph{Topological non-Hermitian origin of surface Maxwell waves}, Nat. Commun. \textbf{10}, 580 (2019).

\bibitem{Stjean2017Lasing} P. St-Jean, V. Goblot, E. Galopin, A. Lema\^{i}tre, T. Ozawa, L. Le Gratiet, I. Sagnes, J. Bloch  and A. Amo, \emph{Lasing in topological edge states of a one-dimensional lattice}, Nat. Photonics \textbf{11}, 651 (2017).

\bibitem{Bahari2017Nonreciprocal} B. Bahari, A. Ndao, F. Vallini, A. E. Amili, Y. Fainman and B. Kant\'{e}, \emph{Nonreciprocal lasing in topological cavities of arbitrary geometries}, Science \textbf{358}, 636 (2017).

\bibitem{Harari2018Topological} G. Harari, M. A. Bandres, Y. Lumer, M. C. Rechtsman, Y. D. Chong, M. Khajavikhan, D. N. Christodoulides and M. Segev, \emph{Topological insulator laser: Theory}, Science \textbf{359}, eaar4003 (2018).

\bibitem{Bandres2018Topological} M. A. Bandres, S. Wittek, G. Harari, M. Parto, J. Ren, M. Segev, D. N. Christodoulides and M. Khajavikhan, \emph{Topological insulator laser: Experiments}, Science \textbf{359}, eaar4005 (2018).

\bibitem{Perczel2017Topological} J. Perczel, J. Borregaard, D. E. Chang, H. Pichler, S. F. Yelin, P. Zoller and M. D. Lukin, \emph{Topological Quantum Optics in Two-Dimensional Atomic Arrays}, Phys. Rev. Lett. \textbf{119}, 023603 (2017).

\bibitem{Pan2015} J.-S. Pan, X.-J. Liu, W. Zhang, W. Yi and G.-C. Guo, \emph{Topological Superradiant States in a Degenerate Fermi Gas}, Phys. Rev. Lett. \textbf{115}, 045303 (2015).

\bibitem{Mivehvar2017} F. Mivehvar, H. Ritsch and F. Piazza, \emph{Superradiant Topological Peierls Insulator inside an Optical Cavity}, Phys. Rev. Lett. \textbf{118}, 073602 (2017).

\bibitem{PhysRevLett.119.173901} P. Doyeux, S. A. Hassani Gangaraj, G. W. Hanson and M. Antezza, \emph{Giant Interatomic Energy-Transport Amplification with Nonreciprocal Photonic Topological Insulators}, Phys. Rev. Lett. \textbf{119}, 173901 (2017).

\bibitem{2017arXiv171100478B} S. Barik, A. Karasahin, C. Flower, T. Cai, H. Miyake, W. DeGottardi, M. Hafezi and E. Waks, \emph{A topological quantum optics interface}, Science \textbf{359}, 666 (2018).

\bibitem{PhysRev.93.99} R. H. Dicke, \emph{Coherence in Spontaneous Radiation Processes}, Phys. Rev. \textbf{93}, 99 (1954).

\bibitem{PhysRevLett.102.143601} M. O. Scully, \emph{Collective Lamb Shift in Single Photon Dicke Superradiance}, Phys. Rev. Lett. \textbf{102}, 143601 (2009).

\bibitem{PhysRevLett.115.243602} M. O. Scully, \emph{Single Photon Subradiance: Quantum Control of Spontaneous Emission and Ultrafast Readout}, Phys. Rev. Lett. \textbf{115}, 243602 (2015).

\bibitem{PhysRevLett.116.163604} P. Tighineanu, R. S. Daveau, T. B. Lehmann, H. E. Beere, D. A. Ritchie, P. Lodahl and S. Stobbe, \emph{Single-Photon Superradiance from a Quantum Dot}, Phys. Rev. Lett. \textbf{116}, 163604 (2016).

\bibitem{PhysRevLett.117.073003} S. J. Roof, K. J. Kemp, M. D. Havey and I. M. Sokolov, \emph{Observation of Single-Photon Superradiance and the Cooperative Lamb Shift in an Extended Sample of Cold Atoms}, Phys. Rev. Lett. \textbf{117}, 073003 (2016).

\bibitem{PhysRevLett.120.193601} L. Chen, P. Wang, Z. Meng, L. Huang, H. Cai, D.-W. Wang, S.-Y. Zhu and J. Zhang, \emph{Experimental Observation of One-Dimensional Superradiance Lattices in Ultracold Atoms}, Phys. Rev. Lett. \textbf{120}, 193601 (2018).

\bibitem{vuletic2006quantum} V. Vuletic, \emph{When Superatoms Talk Photons}, Nat. Phys. \textbf{2}, 801 (2006).

\bibitem{PhysRevLett.99.163601} R. Heidemann, U. Raitzsch, V. Bendkowsky, B. Butscher, R. L\"{o}w, L. Santos and T. Pfau, \emph{Evidence for Coherent Collective Rydberg Excitation in the Strong Blockade Regime}, Phys. Rev. Lett. \textbf{99}, 163601 (2007).

\bibitem{PhysRevX.7.041010} A. Paris-Mandoki, C. Braun, J. Kumlin, C. Tresp, I. Mirgorodskiy, F. Christaller, H. P. B\"{u}chler and S. Hofferberth, \emph{Free-Space Quantum Electrodynamics with a Single Rydberg Superatom}, Phys. Rev. X \textbf{7}, 041010 (2017).

\bibitem{Bahri2015Localization} Y. Bahri, R. Vosk, E. Altman and A. Vishwanath, \emph{Localization and topology protected quantum coherence at the edge of hot matter}, Nat. Commun. \textbf{6}, 7341 (2015).

\bibitem{PhysRevA.96.053858} L. Campos Venuti, Z. Ma, H. Saleur and S. Haas, \emph{Topological protection of coherence in a dissipative environment}, Phys. Rev. A \textbf{96}, 053858 (2017).

\bibitem{Kemp2017Long} J. Kemp, N. Y. Yao, C. R. Laumann and P. Fendley, \emph{Long coherence times for edge spins}, J. Stat. Mech. \textbf{2017}, 063105 (2017).

\bibitem{PhysRevLett.119.123601} I.-D. Potirniche, A. C. Potter, M. Schleier-Smith, A. Vishwanath, and N. Y. Yao, \emph{Floquet Symmetry-Protected Topological Phases in Cold-Atom Systems}, Phys. Rev. Lett. \textbf{119}, 123601 (2017).

\bibitem{PhysRevLett.99.020503} P. Milman, W. Maineult, S. Guibal, L. Guidoni, B. Dou\c{c}ot, L. Ioffe, and T. Coudreau, \emph{Topologically Decoherence-Protected Qubits with Trapped Ions}, Phys. Rev. Lett. \textbf{99}, 020503 (2007).

\bibitem{RevModPhys.80.1083} C. Nayak, S. H. Simon, A. Stern, M. Freedman, and S. D. Sarma, \emph{Non-Abelian anyons and topological quantum computation}, Rev. Mod. Phys. \textbf{80}, 1083 (2008).

\bibitem{Diehl2011Topology} S. Diehl, E. Rico, M. A. Baranov and P. Zoller, \emph{Topology by dissipation in atomic quantum wires}, Nat. Phys. \textbf{7}, 971 (2011).

\bibitem{PhysRevA.79.040303} Z.-Y. Xue, S.-L. Zhu, J. Q. You and Z. D. Wang, \emph{Implementing topological quantum manipulation with superconducting circuits}, Phys. Rev. A \textbf{79}, 040303(R) (2009).

\bibitem{PhysRevLett.109.257002} M. Trif and Y. Tserkovnyak, \emph{Resonantly Tunable Majorana Polariton in a Microwave Cavity}, Phys. Rev. Lett. \textbf{109}, 257002 (2012).

\bibitem{PhysRevLett.110.107006} T. L. Schmidt, A. Nunnenkamp and C. Bruder, \emph{Majorana Qubit Rotations in Microwave Cavities}, Phys. Rev. Lett. \textbf{110}, 107006 (2013).

\bibitem{PhysRevLett.118.126803} M. C. Dartiailh, T. Kontos, B. Dou\c{c}ot and A. Cottet, \emph{Direct Cavity Detection of Majorana Pairs}, Phys. Rev. Lett. \textbf{118}, 126803 (2017).

\bibitem{Konig2007} M. K\"{o}nig, S. Wiedmann, C. Br\"{u}ne, A. Roth, H. Buhmann, L. W. Molenkamp, X.-L. Qi and  S.-C. Zhang, \emph{Quantum Spin Hall Insulator State in HgTe Quantum Wells}, Science \textbf{318}, 766 (2007).

\bibitem{Hsieh2009} D. Hsieh, Y. Xia, D. Qian, L. Wray, J. H. Dil, F. Meier, J. Osterwalder, L. Patthey, J. G. Checkelsky, N. P. Ong, A. V. Fedorov, H. Lin, A. Bansil, D. Grauer, Y. S. Hor, R. J. Cava and M. Z. Hasan, \emph{A tunable topological insulator in the spin helical Dirac transport regime}, Nature \textbf{460}, 1101 (2009).

\bibitem{Wang2009} Z. Wang, Y. Chong, J. D. Joannopoulos and M. Solja\u{c}i\'{c}, \emph{Observation of unidirectional backscattering-immune topological electromagnetic states}, Nature \textbf{461}, 772 (2009).

\bibitem{Hafezi2011} M. Hafezi, E. A. Demler, M. D. Lukin and J. M. Taylor, \emph{Robust optical delay lines with topological protection}, Nat. Phys. \textbf{7}, 907 (2011).

\bibitem{Chang2013} C.-Z. Chang, J. Zhang, X. Feng, J. Shen, Z. Zhang, M. Guo, K. Li, Y. Ou, P. Wei, L.-L. Wang, Z.-Q. Ji, Y. Feng, S. Ji, X. Chen, J. Jia, X. Dai, Z. Fang, S.-C. Zhang, K. He, Y. Wang, L. Lu, X.-C. Ma and Q.-K. Xue, \emph{Experimental Observation of the Quantum Anomalous Hall Effect in a Magnetic Topological Insulator}, Science \textbf{340}, 167 (2013).

\bibitem{Buluta} I. Buluta, S. Ashhab, and F. Nori, \emph{Natural and artificial atoms for quantum computation}, Rep. Prog. Phys. \textbf{74}, 104401 (2011).

\bibitem{Gu2017Microwave} X. Gu, A. F. Kockum, A. Miranowicz, Y.-X. Liu and F. Nori, \emph{Microwave photonics with superconducting quantum circuits}, Phys. Rep. \textbf{718-719}, 1 (2017).

\bibitem{PhysRevApplied.6.044010} J. H. B\'ejanin, T. G. McConkey, J. R. Rinehart, C. T. Earnest, C. R. H. McRae, D. Shiri, J. D. Bateman, Y. Rohanizadegan, B. Penava, P. Breul, S. Royak, M. Zapatka, A. G. Fowler and M. Mariantoni, \emph{Three-Dimensional Wiring for Extensible Quantum Computing: The Quantum Socket}, Phys. Rev. Applied \textbf{6}, 044010 (2016).

\bibitem{QiangLiu2017} Q. Liu, M. Li, K. Dai, K. Zhang, G. Xue, X. Tan, H. Yu and Y. Yu, \emph{Extensible 3D architecture for superconducting quantum computing}, Appl. Phys. Lett. \textbf{110}, 232602 (2017).

\bibitem{Rosenberg2017} D. Rosenberg, D. Kim, R. Das, D. Yost, S. Gustavsson, D. Hover, P. Krantz, A. Melville, L. Racz, G. O. Samach, S. J. Weber, F. Yan, J. L. Yoder, A. J. Kerman and W. D. Oliver, \emph{3D integrated superconducting qubits}, npj Quantum Information \textbf{3}, 42 (2017).

\bibitem{Dunsworth2018} A. Dunsworth, R. Barends, Y. Chen, Z. Chen, B. Chiaro, A. Fowler, B. Foxen, E. Jeffrey, J. Kelly, P. V. Klimov, E. Lucero, J. Y. Mutus, M. Neeley, C. Neil, C. Quintana, P. Roushan, D. Sank, A. Vainsencher, J. Wenner, T. C. White, H. Neven, J. M. Martinis and A. Megrant, \emph{A method for building low loss multi-layer wiring for superconducting microwave devices}, Appl. Phys. Lett. \textbf{112}, 063502 (2018).

\bibitem{Chen2014}  Z. Chen, A. Megrant, J. Kelly, R. Barends, J. Bochmann, Y. Chen, B. Chiaro, A. Dunsworth, E. Jeffrey, J. Y. Mutus, P. J. J. O'Malley, C. Neill, P. Roushan, D. Sank, A. Vainsencher, J. Wenner, T. C. White, A. N. Cleland and J. M. Martinis, \emph{Fabrication and characterization of aluminum airbridges for superconducting microwave circuits}, Appl. Phys. Lett. \textbf{104}, 052602 (2014).

\bibitem{Mukai2019} H. Mukai, K. Sakata, S. J. Devitt, R. Wang, Y. Zhou, Y. Nakajima and J. S. Tsai, \emph{Pseudo-2D superconducting quantum computing circuit for the surface code}, arXiv:1902.07911 (2019).

\bibitem{Rosenberg2019} D. Rosenberg, S. Weber, D. Conway, D. Yost, J. Mallek, G. Calusine, R. Das, D. Kim, M. Schwartz, W. Woods, J. L. Yoder and W. D. Oliver, \emph{3D integration and packaging for solid-state qubits}, arXiv:1906.11146 (2019).

\bibitem{PhysRevLett.85.2392} S.-B. Zheng and G.-C. Guo, \emph{Efficient Scheme for Two-Atom Entanglement and Quantum Information Processing in Cavity QED}, Phys. Rev. Lett. \textbf{85}, 2392 (2000).

\bibitem{PhysRevLett.87.037902} S. Osnaghi, P. Bertet, A. Auffeves, P. Maioli, M. Brune, J. M. Raimond and S. Haroche, \emph{Coherent Control of an Atomic Collision in a Cavity}, Phys. Rev. Lett. \textbf{87}, 037902 (2001).

\bibitem{Majer2007} J. Majer, J. M. Chow, J. M. Gambetta, J. Koch, B. R. Johnson, J. A. Schreier, L. Frunzio, D. I. Schuster, A. A. Houck, A. Wallraff, A. Blais, M. H. Devoret, S. M. Girvin and R. J. Schoelkopf, \emph{Coupling superconducting qubits via a cavity bus}, Nature \textbf{449}, 443 (2007).

\bibitem{Evans2018} R. E. Evans, M. K. Bhaskar, D. D. Sukachev, C. T. Nguyen, A. Sipahigil, M. J. Burek, B. Machielse, G. H. Zhang, A. S. Zibrov, E. Bielejec, H. Park, M. Lon\v{c}ar and M. D. Lukin, \emph{Photon-mediated interactions between quantum emitters in a diamond nanocavity}, Nature \textbf{362}, 662 (2018).

\bibitem{SupplementalMaterials} See Supplemental Material online for additional details about 3D integrated superconducting quantum circuits, effective Hamiltonian, edge states, driving the topological superatom in a cavity, and the effects of symmetry breaking and disorder.

\bibitem{PhysRevX.8.011002} V. D. Vaidya, Y. Guo, R. M. Kroeze, K. E. Ballantine, A. J. Koll\'{a}r, J. Keeling and B. L. Lev, \emph{Tunable-Range, Photon-Mediated Atomic Interactions in Multimode Cavity QED}, Phys. Rev. X  \textbf{8}, 011002 (2018).

\bibitem{PhysRevLett.120.050507} K. Xu, J.-J. Chen, Y. Zeng, Y.-R. Zhang, C. Song, W. Liu, Q. Guo, P. Zhang, D. Xu, H. Deng, K. Huang, H. Wang, X. Zhu, D. Zheng and H. Fan, \emph{Emulating Many-Body Localization with a Superconducting Quantum Processor}, Phys. Rev. Lett. \textbf{120}, 050507 (2018).

\bibitem{Norcia2018} M. A. Norcia, R. J. Lewis-Swan, J. R. K. Cline, B. Zhu, A. M. Rey and J. K. Thompson, \emph{Cavity-mediated collective spin-exchange interactions in a strontium superradiant laser}, Science \textbf{361}, 259 (2018).

\bibitem{ryu2002} S. Ryu and Y. Hatsugai, \emph{Topological Origin of Zero-Energy Edge States in Particle-Hole Symmetric Systems}, Phys. Rev. Lett. \textbf{89}, 077002 (2002).

\bibitem{PhysRevB.78.195125} A. P. Schnyder, S. Ryu, A. Furusaki and A. W. W. Ludwig, \emph{Classification of topological insulators and superconductors in three spatial dimensions}, Phys. Rev. B \textbf{78}, 195125 (2008).

\bibitem{PhysRevLett.115.177204} G. Zhang and Z. Song, \emph{Topological Characterization of Extended Quantum Ising Models}, Phys. Rev. Lett. \textbf{115}, 177204 (2015).

\bibitem{TaoLiu2018} T. Liu, Y.-R. Zhang, Q. Ai, Z. Gong, K. Kawabata, M. Ueda and F. Nori, \emph{Second-Order Topological Phases in Non-Hermitian Systems}, Phys. Rev. Lett. \textbf{122}, 076801  (2019).

\bibitem{Konig08} M. K\"{o}nig, H. Buhmann, L. W. Molenkamp, T. Hughes, C.-X. Liu, X.-L. Qi and S.-C. Zhang, \emph{The Quantum Spin Hall Effect: Theory and Experiment}, J. Phys. Soc. Jpn. \textbf{77}, 031007 (2008).

\bibitem{PhysRevB.92.020515} \'{E}. Dumur, B. K\"{u}ng, A. K. Feofanov, T. Weissl, N. Roch, C. Naud, W. Guichard and O. Buisson, \emph{V-shaped superconducting artificial atom based on two inductively coupled transmons}, Phys. Rev. B \textbf{92}, 020515 (2015).

\bibitem{PhysRevA.94.033829} P.-O. Guimond, H. Pichler, A. Rauschenbeutel and P. Zoller, \emph{Chiral quantum optics with V-level atoms and coherent quantum feedback}, Phys. Rev. A \textbf{94}, 033829 (2016).

\bibitem{Nathan2018} N. Shammah, S. Ahmed, N. Lambert, S. D. Liberato and F. Nori, \emph{Open quantum systems with local and collective incoherent processes: Efficient numerical simulations using permutational invariance}, Phys. Rev. A \textbf{98}, 063815 (2018).

\bibitem{Agarwal1974} G.~S. Agarwal, {\em Quantum Statistical Theories of Spontaneous Emission and their Relation to Other Approaches}, Vol.~70 of {\em Springer tracts in modern physics}, G. H{\"o}hler (Ed.) (Springer, Berlin, 1974).

\bibitem{PhysRevLett.77.3995} P. Zhou and S. Swain, \emph{Ultranarrow Spectral Lines via Quantum Interference}, Phys. Rev. Lett. \textbf{77}, 3995 (1996).

\bibitem{PhysRevLett.76.388} S.-Y. Zhu and M. O. Scully, \emph{Spectral Line Elimination and Spontaneous Emission Cancellation via Quantum Interference}, Phys. Rev. Lett. \textbf{76}, 388 (1996).

\bibitem{PhysRevLett.84.5500} G. S. Agarwal, \emph{Anisotropic Vacuum-Induced Interference in Decay Channels}, Phys. Rev. Lett. \textbf{84}, 5500 (2000).

\bibitem{PhysRevLett.100.043601} Y. P. Yang, J. P. Xu, and S. Y. Zhu, \emph{Quantum Interference Enhancement with Left-Handed Materials}, Phys. Rev. Lett. \textbf{100}, 043601 (2008).

\bibitem{PhysRevLett.101.153601} S. Das, G. S. Agarwal, and M. O. Scully, \emph{Quantum Interferences in Cooperative Dicke Emission from Spatial Variation of the Laser Phase}, Phys. Rev. Lett. \textbf{101}, 153601 (2008).

\bibitem{PhysRevLett.106.020501} A. Gonzalez-Tudela, D. Martin-Cano, E. Moreno, L. Martin-Moreno, C. Tejedor and F. J. Garcia-Vidal, \emph{Entanglement of Two Qubits Mediated by One-Dimensional Plasmonic Waveguides}, Phys. Rev. Lett. \textbf{106}, 020501 (2011).

\bibitem{vanLoo2013} A. F. van Loo, A. Fedorov, K. Lalumi\`{e}re, B. C. Sanders, A. Blais and A. Wallraff, \emph{Photon-Mediated Interactions Between Distant Artificial Atoms}, Science \textbf{342}, 1494 (2013).

\bibitem{Mohammad2018} M. Mirhosseini, E. Kim, X. Zhang, A. Sipahigil, P. B. Dieterle, A. J. Keller, A. Asenjo-Garcia, D. E. Chang and O. Painter, \emph{Cavity quantum electrodynamics with atom-like mirrors}, Nature \textbf{569}, 692 (2019).

\bibitem{McKeever2003} J. McKeever, A. Boca, A. D. Boozer, J. R. Buck and H. J. Kimble, \emph{Experimental realization of a one-atom laser in the regime of strong coupling}, Nature \textbf{425}, 268 (2003).

\bibitem{Astafiev2010Resonance} O. Astafiev, A. M. Zagoskin, A. A. Abdumalikov Jr., Y. A. Pashkin, T. Yamamoto, K. Inomata, Y. Nakamura and J. S. Tsai, \emph{Resonance Fluorescence of a Single Artificial Atom}, Science \textbf{327}, 840 (2010).

\bibitem{PhysRevA.81.033622} X.-J. Liu, X. Liu, C. Wu and J. Sinova, \emph{Quantum anomalous Hall effect with cold atoms trapped in a square lattice}, Phys. Rev. A \textbf{81}, 033622 (2010).

\bibitem{Rechtsmann2013} M. C. Rechtsmann, J. M. Zeuner, Y. Plotnik, Y. Lumer, D. Podolsky, F. Dreisow, S. Nolte, M. Segev and A. Szameit, \emph{Photonic Floquet topological insulators}, Nature \textbf{496}, 196 (2013).

\bibitem{xiao2014} M. Xiao, Z. Q. Zhang and C. T. Chan, \emph{Surface Impedance and Bulk Band Geometric Phases in One-Dimensional Systems}, Phys. Rev. X \textbf{4}, 021017 (2014).

\bibitem{mei2015} F. Mei, J.-B. You, W. Nie, R. Fazio, S.-L. Zhu and L. C. Kwek, \emph{Simulation and detection of photonic Chern insulators in a one-dimensional circuit-QED lattice}, Phys. Rev. A \textbf{92}, 041805(R) (2015).

\bibitem{PhysRevLett.117.210503} K. Kakuyanagi, Y. Matsuzaki, C. D\'{e}prez, H. Toida, K. Semba, H. Yamaguchi, W. J. Munro and S. Saito, \emph{Observation of Collective Coupling between an Engineered Ensemble of Macroscopic
Artificial Atoms and a Superconducting Resonator}, Phys. Rev. Lett. \textbf{117}, 210503 (2016).

\bibitem{PhysRevLett.119.093601} D. Plankensteiner, C. Sommer, H. Ritsch and C. Genes, \emph{Cavity Antiresonance Spectroscopy of Dipole Coupled Subradiant Arrays}, Phys. Rev. Lett. \textbf{119}, 093601 (2017).

\bibitem{Albrecht1} A. Albrecht, L. Henriet, A. Asenjo-Garcia, P. B. Dieterle, O. Painter and D. E. Chang, \emph{Subradiant states of quantum bits coupled to a one-dimensional waveguide}, New J. Phys. \textbf{21} 025003 (2019).

\bibitem{PhysRevLett.121.073602} A. Zhang, K. Zhang, L. Zhou and W. Zhang, \emph{Frozen Condition of Quantum Coherence for Atoms on a Stationary Trajectory}, Phys. Rev. Lett. \textbf{121}, 073602 (2018).

\bibitem{PhysRevLett.123.050502} Y. Ye, Z.-Y. Ge, Y. Wu, S. Wang, M. Gong, Y.-R. Zhang, Q. Zhu, R. Yang, S. Li, F. Liang, J. Lin, Y. Xu, C. Guo, L. Sun, C. Cheng, N. Ma, Z. Y. Meng, H. Deng, H. Rong, C.-Y. Lu, C.-Z. Peng, H. Fan, X. Zhu and J.-W. Pan, \emph{Propagation and Localization of Collective Excitations on a 24-Qubit Superconducting Processor}, Phys. Rev. Lett. \textbf{123}, 050502 (2019).


\bibitem{PhysRevLett.119.250401} L. Pezz\`{e}, M. Gabbrielli, L. Lepori, and A. Smerzi, \emph{Multipartite Entanglement in Topological Quantum Phases}, Phys. Rev. Lett. \textbf{119}, 250401 (2017).

\bibitem{PhysRevLett.120.250501} Y.-R. Zhang, Y. Zeng, H. Fan, J. Q. You and F. Nori, \emph{Characterization of Topological States via Dual Multipartite Entanglement}, Phys. Rev. Lett. \textbf{120}, 250501 (2018).

\end{thebibliography}

\begin{thebibliography}{99}

\bibitem[S1]{PhysRevApplied.6.044010} J. H. B\'ejanin, T. G. McConkey, J. R. Rinehart, C. T. Earnest, C. R. H. McRae, D. Shiri, J. D. Bateman, Y. Rohanizadegan, B. Penava, P. Breul, S. Royak, M. Zapatka, A. G. Fowler and M. Mariantoni, \emph{Three-Dimensional Wiring for Extensible Quantum Computing: The Quantum Socket}, Phys. Rev. Applied \textbf{6}, 044010 (2016).

\bibitem[S2]{Rosenberg2017SM} D. Rosenberg, D. Kim, R. Das, D. Yost, S. Gustavsson, D. Hover, P. Krantz, A. Melville, L. Racz, G. O. Samach, S. J. Weber, F. Yan, J. L. Yoder, A. J. Kerman and W. D. Oliver, \emph{3D integrated superconducting qubits}, npj Quantum Information \textbf{3}, 42 (2017).

\bibitem[S3]{Dunsworth2018SM} A. Dunsworth, R. Barends, Y. Chen, Z. Chen, B. Chiaro, A. Fowler, B. Foxen, E. Jeffrey, J. Kelly, P. V. Klimov, E. Lucero, J. Y. Mutus, M. Neeley, C. Neil, C. Quintana, P. Roushan, D. Sank, A. Vainsencher, J. Wenner, T. C. White, H. Neven, J. M. Martinis and A. Megrant, \emph{A method for building low loss multi-layer wiring for superconducting microwave devices}, Appl. Phys. Lett. \textbf{112}, 063502 (2018).

\bibitem[S4]{Mukai2019SM} H. Mukai, K. Sakata, S. J. Devitt, R. Wang, Y. Zhou, Y. Nakajima and J. S. Tsai, \emph{Pseudo-2D superconducting quantum computing circuit for the surface code}, arXiv:1902.07911 (2019).

\bibitem[S5]{Tudela2011} A. Gonzalez-Tudela, D. Martin-Cano, E. Moreno, L. Martin-Moreno, C. Tejedor and F. J. Garcia-Vidal, \emph{Entanglement of Two Qubits Mediated by One-Dimensional Plasmonic Waveguides}, Phys. Rev. Lett. \textbf{106}, 020501 (2011).

\bibitem[S6]{Mohammad} M. Mirhosseini, E. Kim, X. Zhang, A. Sipahigil, P. B. Dieterle, A. J. Keller, A. Asenjo-Garcia, D. E. Chang and O. Painter, \emph{Cavity quantum electrodynamics with atom-like mirrors}, Nature \textbf{569}, 692 (2019).

\bibitem[S7]{Konig2008} M. K\"{o}nig, H. Buhmann, L. W. Molenkamp, T. Hughes, C.-X. Liu, X.-L. Qi and S.-C. Zhang, \emph{The Quantum Spin Hall Effect: Theory and Experiment}, J. Phys. Soc. Jpn. \textbf{77}, 031007 (2008).

\bibitem[S8]{PhysRevA.81.033622S} X.-J. Liu, X. Liu, C. Wu and J. Sinova, \emph{Quantum anomalous Hall effect with cold atoms trapped in a square lattice}, Phys. Rev. A \textbf{81}, 033622 (2010).

\end{thebibliography}
\end{document}